\documentclass{aa}
\usepackage[varg]{txfonts}
\usepackage{graphicx}
\usepackage{amsmath}
\usepackage{natbib}
\bibpunct{(}{)}{;}{a}{}{,}

\begin{document}

\title{Atmospheric characterization of Proxima b by coupling the SPHERE high-contrast imager to the ESPRESSO spectrograph} 

\author{C. Lovis\inst{1}
\and I. Snellen\inst{2}
\and D. Mouillet\inst{3,4}
\and F. Pepe\inst{1}
\and F. Wildi\inst{1}
\and N. Astudillo-Defru\inst{1}
\and J.-L. Beuzit\inst{3,4}
\and X. Bonfils\inst{3,4}
\and A.~Cheetham\inst{1}
\and U.~Conod\inst{1}
\and X. Delfosse\inst{3,4}
\and D. Ehrenreich\inst{1}
\and P. Figueira\inst{5}
\and T. Forveille\inst{3,4}
\and J. H. C. Martins\inst{5,6}
\and S.~P.~Quanz\inst{7}
\and N. C.~Santos\inst{5,8}
\and H.-M. Schmid\inst{7}
\and D.~S\'egransan\inst{1}
\and S. Udry\inst{1}
} 

\offprints{C. Lovis, \email{christophe.lovis@unige.ch}}

\institute{Observatoire astronomique de l'Universit\'e de Gen\`eve, 51 ch. des Maillettes, 1290 Versoix, Switzerland
\and Leiden Observatory, Leiden University, Postbus 9513, 2300 RA Leiden, The Netherlands
\and Univ. Grenoble Alpes, IPAG, F-38000 Grenoble, France
\and CNRS, IPAG, F-38000 Grenoble, France
\and Instituto de Astrof\'isica e Ci\^encias do Espa\c co, Universidade do Porto, CAUP, Rua das Estrelas, 4150-762 Porto, Portugal
\and European Southern Observatory, Casilla 19001, Santiago, Chile
\and Institute for Astronomy, ETH Zurich, Wolfgang-Pauli-Strasse 27, 8093 Zurich, Switzerland
\and Departamento de F\'isica e Astronomia, Faculdade de Ci\^encias, Universidade do Porto, Rua do Campo Alegre, 4169-007 Porto, Portugal
}

\date{Received 10 September 2016 / Accepted}

\abstract {The temperate Earth-mass planet Proxima b is the closest exoplanet to Earth and represents what may be our best ever opportunity to search for life outside the Solar System.}
{We aim at directly detecting Proxima b and characterizing its atmosphere by spatially resolving the planet and obtaining high-resolution reflected-light spectra.}
{We propose to develop a coupling interface between the SPHERE high-contrast imager and the new ESPRESSO spectrograph, both installed at ESO VLT. The angular separation of 37 mas between Proxima b and its host star requires the use of visible wavelengths to spatially resolve the planet on a 8.2-m telescope. At an estimated planet-to-star contrast of $\sim10^{-7}$ in reflected light, Proxima b is extremely challenging to detect with SPHERE alone. However, the combination of a $\sim$10$^3$-10$^4$ contrast enhancement from SPHERE to the high spectral resolution of ESPRESSO can reveal the planetary spectral features and disentangle them from the stellar ones.}
{We find that significant but realistic upgrades to SPHERE and ESPRESSO would enable a 5-$\sigma$ detection of the planet and yield a measurement of its true mass and albedo in 20-40 nights of telescope time, assuming an Earth-like atmospheric composition. Moreover, it will be possible to probe the O$_2$ bands at 627, 686 and 760 nm, the water vapour band at 717 nm, and the methane band at 715 nm. In particular, a 3.6-$\sigma$ detection of O$_2$ could be made in about 60 nights of telescope time. Those would need to be spread over 3 years considering optimal observability conditions for the planet.}
{The very existence of Proxima b and the SPHERE-ESPRESSO synergy represent a unique opportunity to detect biosignatures on an exoplanet in the near future. It is also a crucial pathfinder experiment for the development of Extremely Large Telescopes and their instruments, in particular the E-ELT and its high-resolution visible/near-IR spectrograph.}


\titlerunning{Atmospheric characterization of Proxima b}
\authorrunning{C. Lovis et al.}

\maketitle

\section{Introduction}

\subsection{Observing exoplanet atmospheres}

The field of exoplanets has seen tremendous progress over the past two decades, evolving from a niche research field with marginal reputation to mainstream astrophysics. The number of exoplanet properties that have become accessible to observations has been continuously growing. The radial velocity and transit techniques have been the two pillars over which exoplanet studies have developed, providing the two most fundamental physical properties of an exoplanet: mass and radius. Simultaneously, spatially-resolved imaging has been studying the properties of young and massive exoplanets on wide orbits. More recently, the field has been moving towards a more detailed characterization of planets and planetary systems, from their orbital architecture to their internal structure to the composition of their atmospheres. The study of exoplanet atmospheres, in particular, is widely seen as the new frontier in the field, a necessary step to elucidate the nature of the mysterious and ubiquitous super-Earths and mini-Neptunes. It is also the only means of directly addressing the fundamental question: has life evolved on other worlds?

Atmospheric characterization heavily relies on the availability of favourable targets, given the extremely low-amplitude signals to be detected and the present instrumental limitations. That is why a major effort is being made to systematically search for the nearest exoplanets with the largest possible atmospheric signatures. One of the most successful techniques so far has been transit spectroscopy, where an exoplanet atmosphere is illuminated from behind by the host star, and light is transmitted according to the wavelength-dependent opacities of chemical species within the atmosphere. In this approach, the most favourable targets are planets transiting nearby, small stars. Not only gaseous giant planets, but also mini-Neptunes and super-Earths have been probed via transmission spectroscopy, paving the way towards the characterization of truly Earth-like planets. Recently, \citet{Gillon2016} have announced the discovery of a system that is potentially promising in this respect: three transiting Earth-size objects around the 0.08-$M_\odot$ star TRAPPIST-1. More generally, it is expected that the ongoing/upcoming transit searches (TESS, CHEOPS, PLATO, NGTS, MEarth, TRAPPIST, SPECULOOS, ExTrA) will find most of the nearest transiting systems in the next few years, opening a new era of atmospheric characterization with e.g. the James Webb Space Telescope.

An alternative to transit spectroscopy is the detection of exoplanets using high angular resolution, i.e. the capability to spatially resolve the light emitted by an exoplanet from the one of its host star. So far, this technique has been mostly applied to young, self-luminous massive planets on wide orbits, which offer the highest planet-to-star contrasts at the angular separations currently accessible to 10-m class telescopes equipped with state-of-the-art adaptive optics (AO) systems. Given the very challenging flux ratios (10$^{-4}$-10$^{-10}$ at sub-arcsec angular separations), many different technical and observational strategies are being explored to make progress in this field.

A promising avenue for atmospheric characterization is the high-contrast/high-resolution technique (hereafter HCHR). It was first envisaged by \citet{Sparks2002} and \citet{Riaud2007}, and simulated in details by \citet{Snellen2015}. It recently found a first real-life application on the young exoplanet Beta Pic b \citep{Snellen2014}, leading to the detection of CO in the planet thermal spectrum and to the measurement of the planet spin rate. The technique combines a high-contrast AO system to a high-resolution spectrograph to overcome the tiny planet-to-star flux ratio in two steps. In the first stage, the AO system spatially resolves the exoplanet from its host star and enhances the planet-to-star contrast at the planet location. However, the remaining stellar signal (from e.g. the wings of the stellar PSF or non-perfect AO correction) may still be orders of magnitude larger than the planet signal. In the second stage, the light beam at the planet location is sent to a high-resolution spectrograph. Simultaneously, a reference star-only spectrum is recorded from a spatial location away from the planet. The planet spectrum can then be recovered by differencing the two spectra, provided sufficient signal-to-noise ratio (SNR) is achieved. Spectral features that are present both in the planet and star spectrum can be separated thanks to the Doppler shifts induced by the planet orbital velocity. At the same time, measuring the planet orbital velocity enables the measurement of the orbital inclination angle and true mass of the planet, when combined to high-precision RV measurements of the star. 

For the HCHR technique to be applicable, it is necessary to spatially resolve the star-planet system. The distance to the star therefore plays a critical role. Finding exoplanets orbiting the very nearest stars (within $\sim$5-10 pc) is thus a necessary step for the future of this technique, and atmospheric characterization in general.

\subsection{Proxima b}
\label{SectProxima}

\citet{Anglada2016} have recently announced the detection of a low-mass exoplanet candidate around our closest stellar neighbour, Proxima Centauri ($d$ = 1.30 pc). Proxima is a 0.12-$M_\odot$ red dwarf (spectral type M5.5V) with a bolometric luminosity of 0.00155 $L_\odot$ \citep{Boyajian2012}, which translates into a visual magnitude of only $V$ = 11.13 despite the proximity of the star. The planet, Proxima b, has an orbital period of 11.2 days and a semi-major axis of 0.048 au. Its minimum mass is 1.27 $\pm$ 0.18 $M_\oplus$. The planet discovery is based on long-term, high-cadence radial velocity (RV) time series obtained with the HARPS and UVES spectrographs \citep{Mayor2003,Dekker2000} at the European Southern Observatory. Proxima b is exceptional in several respects: it is the closest exoplanet to Earth there will ever be; with a minimum mass of 1.3 $M_\oplus$ it is likely to be rocky in composition; and with a stellar irradiation that is about 67\% of Earth's irradiation it is plausible that habitable conditions prevail at least in some regions of the planet. This is especially true if the planet has a fairly thick atmosphere, as could be expected from its mass that is larger than Earth.

The question of a habitable climate on Proxima b is a complex one that is being addressed by numerous studies \citep[e.g.][]{Ribas2016,Turbet2016,Meadows2016}. The spin rate of the planet is likely to be either in a tidally-locked state or in a 3:2 orbital resonance, critically impacting global atmospheric circulation and climate \citep[although the atmosphere itself may force the spin rate into a different regime as in Venus, e.g.][]{Leconte2015}. If the planet was formed {\it in situ}, it likely experienced a runaway greenhouse phase causing a partial loss of water with some uncertainties about the quantity of water which may subsist \citep{Ribas2016,Barnes2016}. If water is still present today, the amount of greenhouse gases in the atmosphere can drive the system into very different states: from a temperate, habitable climate to a cold trap on the night side causing atmospheric collapse. Adding to the complexity, the high X-ray and UV flux of the star may have caused atmospheric evaporation that could have a major effect on the planet volatile content, in particular water. Note however that the initial water content itself is not constrained; if the planet was formed beyond the ice line and migrated inwards, it could have been born as a water world and have remained so until today. Despite all these uncertainties, Proxima b concentrates on itself a number of properties that make it a landmark discovery.

The immediate question that comes to mind is how to study this outstanding object in more details. With a geometric transit probability of only 1.5\%, the planet is unlikely to transit its host star, strongly limiting the observational means of characterizing it. The James Webb Space Telescope (launch 2018) may be able to obtain a thermal phase curve of the planet, provided its day-to-night temperature contrast is sufficiently high \citep{Kreidberg2016}. In this paper we examine an alternative method that can directly probe Proxima b: the HCHR technique, applied to reflected light from the planet. Quite remarkably, \citet{Snellen2015} anticipated the existence of a planet orbiting Proxima and simulated HCHR observations with the future European Extremely Large Telescope (E-ELT). In the present paper we attempt to go one step further towards a practical implementation, basing our simulations on the real planet Proxima b and existing VLT instrumentation.

Proxima b was discovered with the radial velocity technique. The knowledge of the RV orbit yields two critical pieces of information for the planning and optimization of HCHR observations: the angular separation at maximum elongation from the star (quadrature), and the epochs of maximum elongations. For Proxima b, combining the RV-derived semi-major axis and the known distance to the system, one obtains a maximum angular separation of 37 milli-arcseconds (mas). The RV-derived ephemeris then also provides the timing of the two quadratures within each 11.2-d orbit (we assume here that a regular RV monitoring of the star maintains sufficient accuracy on the ephemerides). What the RV data do not provide is orbital inclination and position angle of the orbit on the sky. Thus, one specific challenge of the HCHR approach is that it first has to find the position of the planet around the star, before being able to study it in details. 

\section{Exoplanets in reflected light}

\subsection{Reflected-light flux ratio}

We start by deriving the properties of exoplanet reflected light spectra from an observer's standpoint. The amount of light that is reflected off a planet towards a distant observer is given by:

\begin{equation}
F_p(\lambda, \alpha) = F_s(\lambda) \, A_g(\lambda) \, g(\alpha) \, \left(\frac{R_p}{a}\right)^2 ,
\label{EqReflectedLight1}
\end{equation}

where $F_p(\lambda, \alpha)$ is the reflected spectrum at phase angle $\alpha$, $F_s(\lambda)$ is the stellar spectrum, $A_g(\lambda)$ is the geometric albedo spectrum, $g(\alpha)$ is the phase function, $R_p$ is the planet radius, and $a$ is the orbital semi-major axis. The phase angle $\alpha$ is defined as $\cos{\alpha} = -\sin{i} \cos{\phi}$, with $i$ the orbital inclination angle relative to the observer, and $\phi$ the orbital phase, defined as zero at inferior conjunction. The phase function $g(\alpha)$ depends on the scattering properties of the planet atmosphere or surface. For simplicity we adopt in this work a Lambert phase function, valid for isotropic scattering:

\begin{equation}
g(\alpha) = \frac{\sin{\alpha} + (\pi - \alpha) \cos{\alpha}}{\pi} .
\end{equation}

Following Eq.~\ref{EqReflectedLight1} one obtains a tiny planet-to-star flux ratio of $1.3 \times 10^{-10}$ for the Earth \citep[$A_g$ $\sim$ 0.22;][]{Robinson2011} at orbital quadrature ($\alpha$ = 90 deg). Interestingly, this ratio increases to $\sim$10$^{-7}$ for an Earth-like planet orbiting a late-M dwarf, thanks to the much lower luminosity of the star and thus closer location of the habitable zone ($<$ 0.1 au). Late-M dwarfs therefore offer a factor of $\sim$100-1000 advantage over solar-type stars for studying habitable planets in reflected light.

At high spectral resolution, Doppler effects must be taken into account in Eq.~\ref{EqReflectedLight1} and introduce a dependence on the planet projected orbital velocity. The expression for the reflected spectrum becomes:

\begin{equation}
F_p(\lambda, v_r, \alpha) = F_s(\lambda, v_r) \, A_g(\lambda, v_r) \, g(\alpha) \, \left(\frac{R_p}{a}\right)^2 ,
\label{EqReflectedLight2}
\end{equation}

where $v_r$ is the projection of the planet orbital velocity onto the line of sight, and the notation $F(\lambda, v_r)$ means that the spectrum $F$ is Doppler-shifted by $v_r$. For simplicity we assume here a circular orbit, so that stellar light hitting the planet is not Doppler-shifted in the planet rest frame (no radial velocity component). We further assume that rotational broadening of the stellar lines is the same for the distant observer as the one seen from the planet.

\subsection{Angular separation}

To evaluate the detectability of an exoplanet with the HCHR technique, the other important aspect besides flux ratio is angular separation from the host star. At orbital quadrature it is simply given by:

\begin{equation}
\theta \, \mathrm{[arcsec]} = \frac{a\, \mathrm{[au]}}{d\, \mathrm{[pc]}} ,
\label{EqAngSep}
\end{equation}

where $d$ is the distance to the system. We note here the critical dependence on $d$, which implies that habitable planets around the nearest M dwarfs are located at angular separations typically $<$ 100 mas. Coincidentally, the diffraction limit of a telescope, $\theta_{\mathrm{diff}} = 1.22\; \lambda/D$, is 30 mas at $\lambda =1.0\;\mu$m for a 8.2-m primary miror. Considering a minimal inner working angle of at least $\sim$2 $\lambda/D$, it follows that habitable worlds need to orbit a very nearby M dwarf ($\lesssim$ 5 pc) to be spatially resolved by a 10-m class telescope. Clearly, the future generation of Extremely Large Telescopes will be necessary to explore these stars beyond just a few parsecs. Habitable planets orbiting higher-mass G and K stars can be probed to larger distances, although at the cost of a much lower contrast.

\begin{figure*}
\centering
\includegraphics[width=15cm]{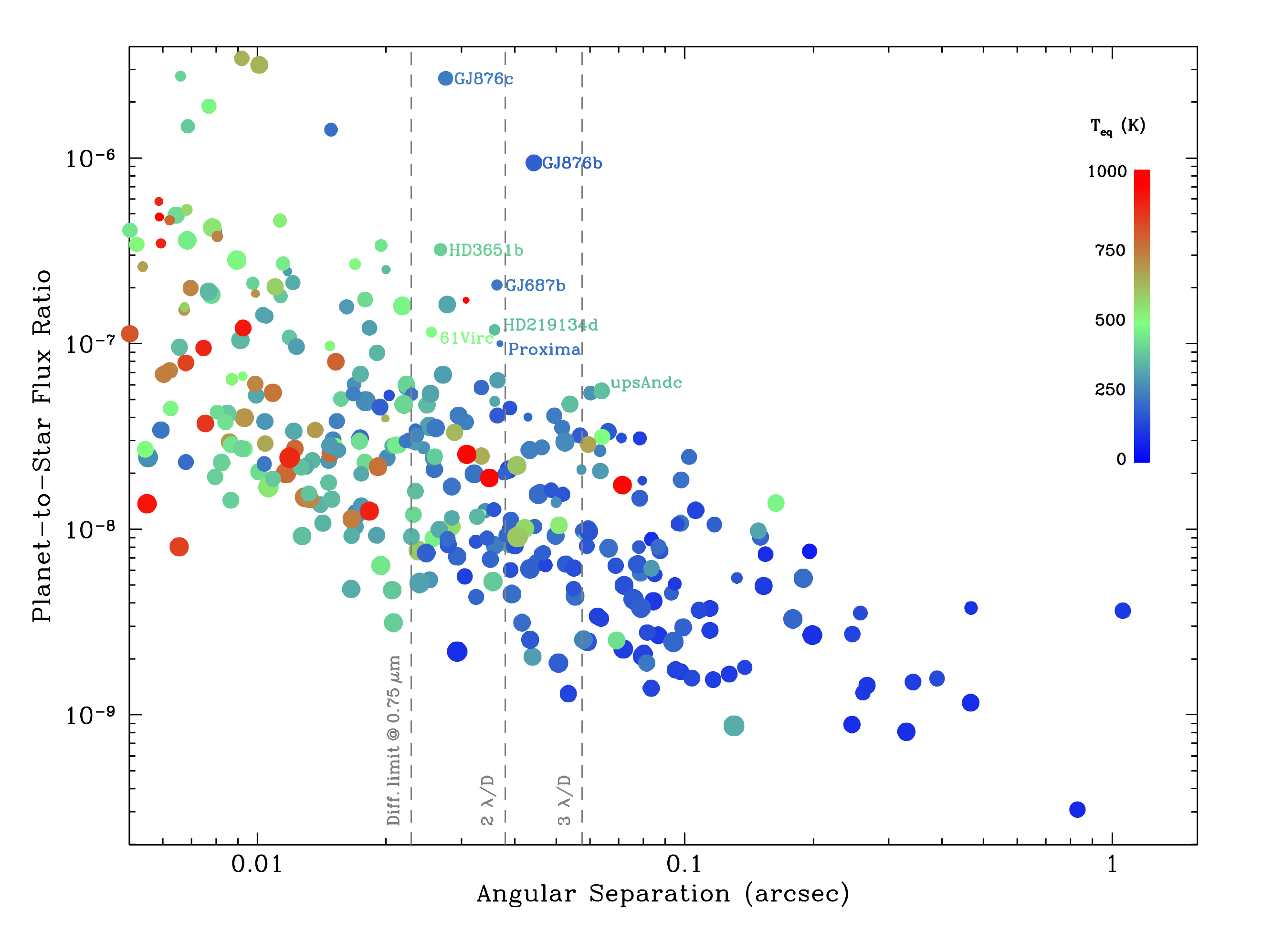}
\caption{Estimated planet-to-star contrast in reflected light for known exoplanets as a function of angular separation from their host star. Dot size is proportional to the logarithm of planet mass, while the color scale represents equilibrium temperature (assuming a Bond albedo of 0.3). Vertical dashed lines indicate the diffraction limit, 2 $\lambda/D$ and 3 $\lambda/D$ thresholds for the 8.2-m VLT at 750 nm (corresponding to the O$_2$ A-band).}
\label{FigReflectedLight}
\end{figure*}

\subsection{Known exoplanets in reflected light}
\label{SectReflectedLight}

Based on Eqs.~\ref{EqReflectedLight2} and \ref{EqAngSep} we proceed to estimate the reflected light contrast for all known exoplanets as a function of their angular separation. The exoplanet list was retrieved from exoplanet.eu as of 30 April 2016, to which we added the newly-discovered Proxima b. Since most nearby exoplanets do not transit, we derive their radius from their minimum mass, by using an average mass-radius relation. We first estimate true masses by multiplying minimum masses by $\sqrt{4/3}$ $\sim$ 1.15, which is the median value of the $(1/\sin{i})$ correction factor in the case of randomly-oriented orbits. Despite many recent advances \citep[e.g.][]{Rogers2015,Dressing2015,Wolfgang2015}, there is no generally-accepted average mass-radius relation for exoplanets. Moreover, it is clear that a diversity of objects co-exist in the super-Earth/mini-Neptune regime, so that even the concept of an average mass-radius relation is questionable. For the present exercise we choose to assign radii based on the piece-wise mass-radius relation derived by \citet{Chen2016}, which describes three different regimes: rocky planets with an Earth-like composition below 2 $M_\oplus$, super-Earths and ice giants between 2-130 $M_\oplus$, and gas giants beyond 130 $M_\oplus$. Obviously, this piece-wise relation is only a rough approximation to the complex reality of exoplanets, as it neglects the intrinsic scatter in composition that occurs at any given mass. This procedure yields an estimated radius of 1.1 $R_\oplus$ for Proxima b (for a derived true mass of 1.5 $M_\oplus$).

Estimating geometric albedos is difficult given the scarce knowledge of exoplanet properties we have today. For the present exercise we use a fixed geometric albedo of 0.40 for all planets, except Proxima b (see below). This value is justified as follows: planets that can be angularly resolved are mostly temperate to cool and are expected to have rather high albedos owing to the presence of reflective clouds \citep[e.g.][]{Cahoy2010}. Finally, the last ingredient needed to compute the flux ratio is the phase function. Here we consider observations around phase $\alpha$ = 90 deg, when the planet is at maximum elongation. The Lambert phase function has value $1/\pi$ in this case, a significant flux drop compared to a full-phase configuration (which is inaccessible if one wants to spatially resolve the planet from its star). We note that the Lambert model is a good approximation for Rayleigh-scattering atmospheres, but may overestimate the flux in other cases, so it should be considered as a best case.

Regarding Proxima b, we benefit from the availability of specific atmospheric models and simply adopt here the flux ratio predicted by \citet{Turbet2016} for the case of an Earth-like atmosphere on a water-rich world (their Fig.~10). It is about 1.0 $\times$ 10$^{-7}$ at a phase angle of 90 deg in the red optical wavelength range. We note however that this number may vary significantly depending on actual atmospheric and surface properties (see Sect.~\ref{SectDetectability} for more details).

The results for all known exoplanets are shown in Fig.~\ref{FigReflectedLight}. The figure shows several interesting features. First, larger angular separations generally mean cooler planets and lower contrast. However, there is a marked diversity of objects at any given location in parameter space due to the large spread in stellar distances and spectral types. In terms of observability with the HCHR technique, an angular separation of $\sim$2 $\lambda/D$ can be considered as a lower limit to spatially resolve the planet. Applying this criterion to the 8.2-m VLT at 750 nm, one can see 4 objects that stand out for being potentially detectable: the cool Jupiter GJ~876~b, the cool and warm Neptunes GJ~687~b and HD~219134~d, and the temperate Earth-mass planet Proxima b. These four planets have an expected contrast at the level of 10$^{-7}$ or higher.

Proxima b appears at a separation of 37 mas from the star. This represents exactly 2.0 $\lambda/D$ at 750 nm, but only 1.2 $\lambda/D$ in $J$-band. It is significantly below the diffraction limit in $H$- and $K$-band. Given that a good AO correction is exceedingly difficult to achieve blueward of the $R$-band, the red optical wavelength range is the only possibility to probe Proxima b (and the other three objects) with the current generation of 10-m class telescopes.

\subsection{Signal-to-noise ratio}
\label{SectSNR}

In the context of the HCHR technique, the signal-to-noise ratio on the planet spectrum can be expressed as \citep{Snellen2015}:

\begin{equation}
SNR_p = \frac{F_p}{\sqrt{F_s/K + \sigma^2_{bkgr} + \sigma^2_{RON} + \sigma^2_{dark}}} \sqrt{N_{\mathrm{lines}}} ,
\label{EqSNR1}
\end{equation}

where $F_p$ and $F_s$ are the measured planet and star fluxes expressed in photo-electrons per high-resolution wavelength bin, $K$ is the AO contrast enhancement factor at the planet location, $\sigma_{bkgr}$, $\sigma_{RON}$ and $\sigma_{dark}$ are the sky background noise, detector readout noise and dark current noise, respectively, and $N_{\mathrm{lines}}$ is the number of spectral lines used in a cross-correlation procedure. Note that $N_{\mathrm{lines}}$, in conjunction with $F_p$, are meant to represent the number and characteristics of typical spectral features in the planetary high-resolution spectrum, in terms of intrinsic width and contrast in particular.

Some clarifications are in order regarding the definition of fluxes and $K$ factor. Eq.~\ref{EqSNR1} considers an observation with the spectrograph slit or fiber centered on the planet. The measured planet flux $F_p$ takes into account the slit coupling efficiency, $\eta_p$. Similarly, $F_s$ is defined as the stellar flux that would be measured with the slit or fiber centered on the star (i.e. including the same coupling efficiency). The actual coupling efficiency of the stellar light {\it at the location of the planet}, which we call $\eta_s$, is much lower than $\eta_p$ and is directly related to the contrast enhancement factor $K$ by $K = \eta_p / \eta_s$. To explicitly show the dependence on the coupling efficiency, we further define $F'_p = F_p/\eta_p$ and $F'_s = F_s/\eta_p$ as the planetary and stellar fluxes that would be detected if the slit efficiency was 100\% {\it and} the planet or star were centered on the slit.

In the case of reflected light, one can insert Eq.~\ref{EqReflectedLight1} into Eq.~\ref{EqSNR1} to obtain:

\begin{equation}
SNR_p = \frac{C\, F'_s\, \eta_p}{\sqrt{F'_s\, \eta_p / K + \sigma^2_{bkgr} + \sigma^2_{RON} + \sigma^2_{dark}}} \sqrt{N_{\mathrm{lines}}} ,
\label{EqSNR2}
\end{equation}

where $C$ = $A_g(\lambda) \, g(\alpha) \, (R_p / a)^2$ is the planet-to-star contrast.

If stellar shot noise dominates the noise budget, Eq.~\ref{EqSNR2} simplifies to:

\begin{equation}
SNR_p = C\, \sqrt{F'_s\, K\, \eta_p\,N_{\mathrm{lines}}} ,
\label{EqSNR3}
\end{equation}

The latter equation (strictly valid only in the photon-limited regime) explicitly shows how SNR is impacted by the instrumental setup, in particular the details of the slit/fiber coupling and AO performance. We will use below Eqs.~\ref{EqSNR2} and \ref{EqSNR3} to design HCHR observations of Proxima b.





\section{The SPHERE-ESPRESSO opportunity}

\subsection{Instrumental context}

Applying the HCHR technique is very demanding in terms of instrumentation. So far it has been demonstrated in only one case: $\beta$ Pictoris b, using the CRIRES infrared spectrograph at the VLT, coupled to the MACAO adaptive optics system \citep{Snellen2014}. In the case of Proxima b we face a much more challenging situation since we must work in the visible regime to resolve the system, where good AO performance is more difficult to achieve.

Fortunately, ESO now offers the SPHERE high-contrast imager \citep{Beuzit2008}, an extreme-AO facility especially designed for high-contrast observations of exoplanets. SPHERE is mounted on the Nasmyth A platform of the VLT UT3 telescope, and comprises three distinct science instruments: IRDIS, IFS, and ZIMPOL. The SPHERE Common Path and Infrastructure (CPI) provides an AO-corrected, coronagraphic light beam to the three instruments. While the first two work in the near-IR, ZIMPOL covers the 500-900 nm wavelength range. In $R$-band, the Strehl ratio reaches more than 40\% on bright stars under good seeing conditions, with a PSF FWHM of order 20 mas \citep[e.g.][]{Kervella2016}.

In parallel, a consortium of European institutions and ESO are presently building the ESPRESSO high-resolution spectrograph, to be installed in the combined Coud\'e laboratory (CCL) of the VLT \citep{Pepe2014}. ESPRESSO can be fed by any of the four UT telescopes (or all at the same time) through four Coud\'e trains that bring the light from the Nasmyth B focus of the telescopes to the CCL. ESPRESSO is a highly-stabilized, fiber-fed, cross-dispersed echelle spectrograph covering the 380-780 nm wavelength range at a spectral resolution $R$ = 130,000 for a 1.0-arcsec sky aperture. Importantly, it will also have an ultra-high resolution mode, $R$ = 220,000, using a 0.5-arcsec aperture. In all modes the spectrograph is fed by two identical fibers conveying the stellar light as well as the sky background light or a spectral reference source. ESPRESSO is expected to be commissioned at Paranal in 2017.

In the context of exoplanet atmosphere studies with the HCHR technique, it is extremely interesting to envisage the possibility to couple SPHERE to ESPRESSO. In fact, ESO VLT is one of the very few 10-m class facilities in the world that offer an {\it extreme} AO system {\it and} a high-resolution spectrograph on the same telescope. We note here that the coupling of SPHERE to another future VLT spectrograph, CRIRES+ covering the near-IR, is already under study (M. Kasper, priv. comm.). However, as shown above, Proxima b cannot be resolved by the VLT at near-IR wavelengths. We thus do not discuss this option further here.

\subsection{Initial search with SPHERE-ZIMPOL}

The ZIMPOL instrument in SPHERE has been specifically designed to achieve extreme polarimetric contrast \citep{Schmid2006}. Fast polarimetric modulation cancels the unpolarized, direct stellar light to a high degree and enables the detection of faint companions in reflected light, provided they exhibit a significant polarization fraction. \citet{Milli2013} provide a detailed assessment of the ZIMPOL expected performances for the detection of exoplanets orbiting nearby stars. They show in particular that ZIMPOL may be able to achieve a polarimetric contrast at the 10$^{-7}$ level close to the star, and estimate that the putative Earth-mass planet Alpha Cen Bb could be detectable if it has favourable polarimetric properties.

Searching for Proxima b with ZIMPOL is a logical first step before attempting the more challenging HCHR approach with ESPRESSO. A detection would yield a measurement of the polarized reflectance of the planet and thus constrain some of its atmospheric properties. It would also pin down the position angle of the planetary orbit on the sky, thereby saving telescope time for subsequent HCHR observations. Overall, a ZIMPOL detection would be highly complementary to HCHR observations, although not mandatory for the latter to be carried out.

\subsection{Implementation of a SPHERE-ESPRESSO coupling}

In practice, and in very broad terms, coupling SPHERE to ESPRESSO would mean building a fiber pick-up interface, possibly as an integral field unit (IFU), in the focal plane of the SPHERE AO-corrected visible beam. One fiber/spaxel would be placed on the expected planet position (once known), while the other ones would collect a star-only reference spectrum. The number of fibers will depend on the ESPRESSO injection capabilities, which we explore further below. The fibers would then convey the light to the ESPRESSO UT3 Coud\'e train optics, or directly to the spectrograph. In this paper we limit ourselves to conceptual ideas regarding modifications and improvements that would be required on the two instruments, and leave the detailed design work to further studies.

The major upgrades that are required compared to the existing instruments are: 1) an enhanced stellar light rejection capability at 37 mas from the star ($\sim$2.0 $\lambda/D$) in the SPHERE focal plane, 2) a fiber pick-up interface within SPHERE, and 3) a dedicated fiber injection channel into ESPRESSO. We discuss these items separately below.

\subsubsection{Optimized AO correction and coronagraph}
\label{SectAO}

\begin{figure*}
\centering
\includegraphics[width=8cm]{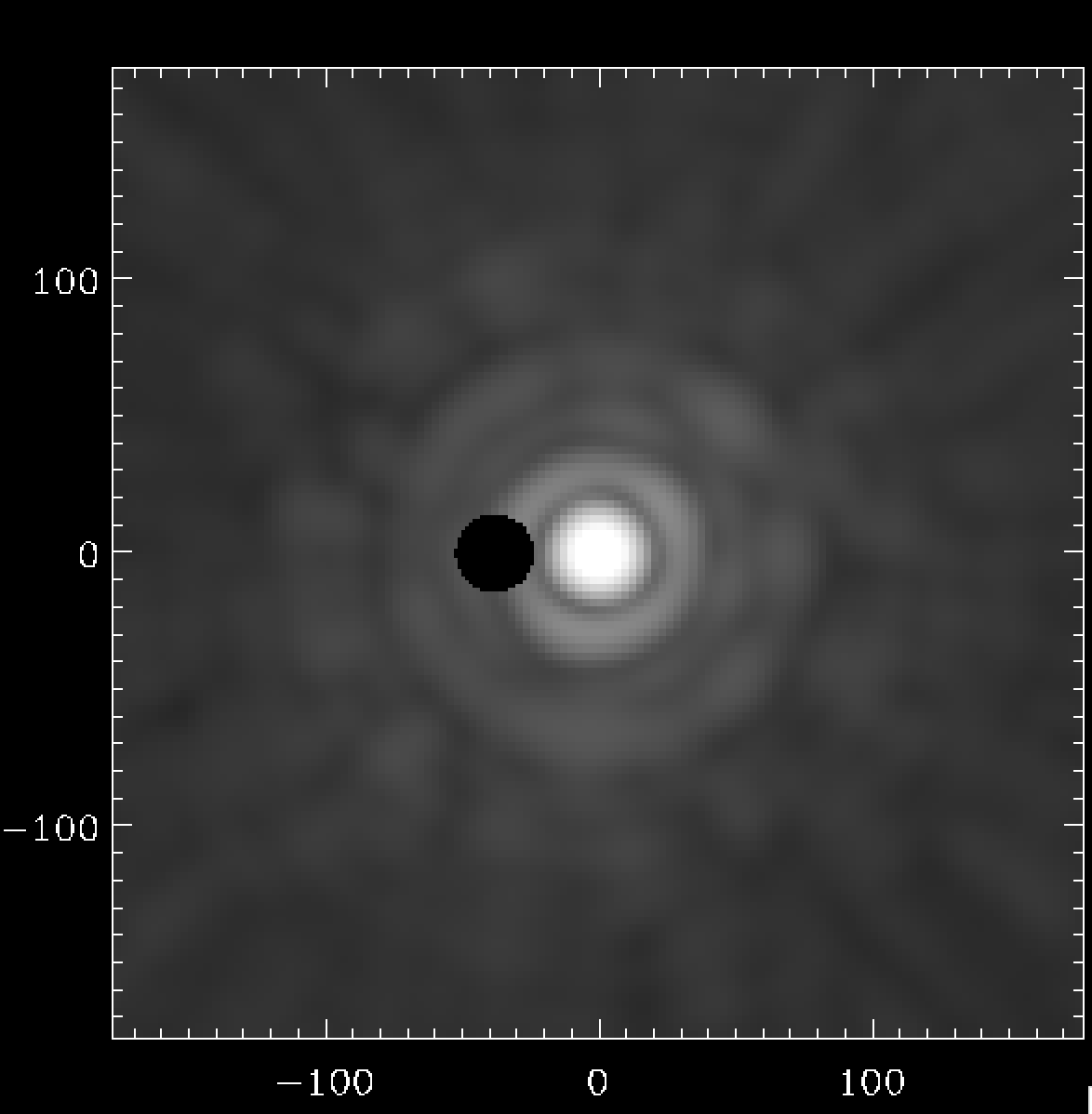}
\includegraphics[width=8cm]{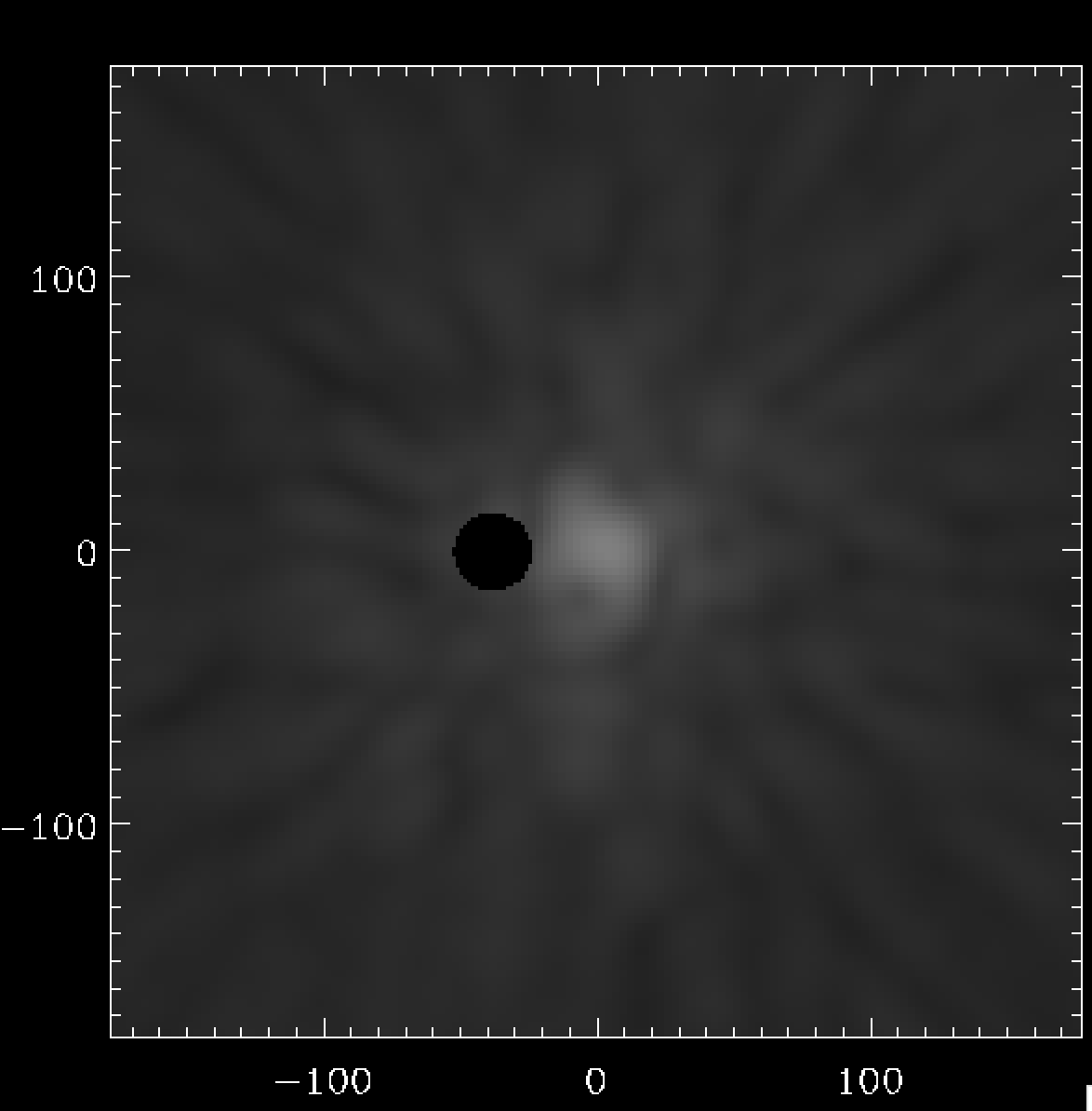}
\caption{Left: realistic simulation of a SPHERE AO-corrected image in the 600-780 nm band, centered on the star. The dark hole indicates the position and size of an optical fiber that would pick up the planet light to send it to ESPRESSO. The planet is 37 mas away from the star (2.0 $\lambda/D$ at 750 nm), while the fiber radius is 14 mas. Right: coronagraphic image, limited by the aberrations of the incoming beam (both turbulence residuals and optical defects), assuming an ideal coronagraph. Units on both axes are milli-arcseconds (mas).}
\label{FigFoV}
\end{figure*}

We first note that Proxima has magnitude $R$ = 9.45 and $I$ = 7.41. SPHERE AO performance has proven to remain nominal up to magnitude $R$ $>$ 11 when using all the visible photons for wavefront sensing \citep{Fusco2016}. When observing in the visible, the photons are shared between the science channel (80\%) and wavefront sensor (20\%), significantly reducing the AO limiting magnitude. For the purpose of this paper, and considering that ESPRESSO does not go beyond 780 nm, we envisage a new dedicated dichroic beamsplitter, providing ESPRESSO (respectively the WFS) with 100\% of the light below (respectively above) 780 nm. This beamsplitter would take less than half of the flux of this red star from the total WFS bandpass. This will keep the AO performance in the bright regime, with achievable Strehl ratio as high as a 50\% in $R$-band under good seeing conditions.

We now attempt to estimate and optimize the level of stellar light rejection that can be achieved with SPHERE. To do this in a realistic way, we use SPHERE AO simulations that are well supported by the image quality demonstrated on sky, under a mean seeing of 0.85 arcsec, regular wind (12 m s$^{-1}$), and a residual jitter < 3 mas. Fig.~\ref{FigFoV} shows a representative image of the SPHERE focal plane under these conditions. We find that a fiber centered on the position of the companion at 37 mas in an AO-corrected non-coronagraphic image suffers from significant contamination by the star diffraction pattern. From the fundamental limit of diffraction, a better rejection of stellar light can only be obtained with a further cancellation effect: coronagraphy, and/or a coherent-light suppression approach. Dedicated coronagraphs can transmit off-axis companions down to very small inner working angles (IWAs). However, before reaching their ultimate intrinsic limitations, they are usually limited in ground-based applications by the aberrations of the incoming beam, especially in the visible with moderate Strehl ratio. This is clearly the case here, so we adopt the generic approach of the so-called ideal coronagraph: the coherent part of the incoming light is cancelled out and the resulting image is directly related to the variance and spatial structure of the aberrations. We assume here a good calibration and centering of the star on the coronagraph to better than 2 mas at the observing wavelength (which requires careful calibration but remains reasonable with SPHERE). Fig.~\ref{FigFoV} shows the expected stellar residuals on the coronagraphic image, given 30 nm of optical aberrations upstream of the coronagraph.

Based on these simulations, we first study which fiber size would be optimal for our observations, i.e. which fiber size maximizes the planet SNR. From Eq.~\ref{EqSNR3} we can define a fiber-coupling figure of merit as $Q = \sqrt{K\, \eta_p}$. We simulate a range of fiber sizes at different wavelengths between 600 and 780 nm, and evaluate $Q$ in each case. We find that a fiber radius of $\sim$14 mas maximizes $Q$ across the wavelength range. At the long end of the range, this radius corresponds to an inner edge of the fiber just off the central Airy disk of the star. From there we directly obtain the values of the coupling parameters $\eta_p$ and $K$. In the non-coronagraphic case, we obtain $\eta_p$ = 38\% and $K$ = 62 (averaged over the wavelength range). For the coronagraphic case, the contrast enhancement factor can be pushed to about $K$ = 530. In this case, $\eta_p$ formally remains the same but the total throughput is reduced because of the non-perfect transmission of the coronagraph. In the following we will assume a coronagraph transmission of 70\%. Overall, the use of a coronagraph enables a gain of a factor $\sim$3 on the planet SNR compared to the non-coronagraphic case. The values reported here correspond to what can be expected in good conditions, careful calibration and setting of SPHERE, with a dedicated setup including a new coronagraph and optimized beamsplitter. We will designate this setup as the "short-term SPHERE upgrade".

Can we consider even better performance? This is certainly difficult and not demonstrated but we are not at the fundamental limit and we can consider further developments, likely to be more accessible and faster to implement than more complex instruments such as those on ELTs. Among further upgrades, we can in particular push further the AO performance. Faster correction would improve the residuals at short separation, which would require to change hardware and software for not only the sensor but also the real-time computer and deformable mirror. Ongoing progress on each of these items make such upgrades possible in the future (and in any case mandatory for the ELTs). Observing in selected observing conditions, in particular under low wind, would also help. Following this path, we will then hit the sensing photon noise limit, but this would also be pushed towards fainter targets using an intrinsically more sensitive wavefront sensor like the pyramid wavefront sensor \citep[e.g.][]{Esposito2011}. Finally, more specifically at the location of the companion, speckle identification (through modulation of the wavefront, and detection of a brightness modulation) and corresponding cancellation can further reduce the stellar contamination. Experimental work and preliminary tests on sky are ongoing. Altogether, an additional improvement by a factor $\sim$10 can be considered at the cost of a significant upgrade of SPHERE, which would however still be easier and faster than building a high-contrast imager on an ELT. For the purpose of this paper, we will assume that a major but still realistic SPHERE upgrade would enable $K$ = 5000 and $\eta_p$ = 60\%. In the following we will designate this setup as the "2nd-generation SPHERE", or simply SPHERE+.

Regarding coronagraphs, we note that the SPHERE visible channel includes various coronagraphs, but none of them is fully appropriate at very small separation: the classical Lyot coronagraphs have a minimum inner working angle (IWA) of 3 $\lambda$/D and the four-quadrant phase mask (4QPM) allows for some companion transmission at smaller separation but with some restrictions on the spectral bandpass and with a blind area in the field of view along the quadrant transition. For the ambitious observations described here, we will need a new, optimized coronagraph. More precisely, we can enumerate a few key properties that such a coronagraph shall have:

\begin{itemize}
\item It should be able to work in the real environment of the SPHERE instrument, including typical residual wavefront errors and image/pupil stability performances
\item The working wavelength range of the coronagraph should be 600-780 nm (i.e. a relative bandwidth of about 25\%)
\item Stellar light rejection should be optimized for angular separations between $\sim$1.0 and 3.0 $\lambda/D$
\item The inner working angle (IWA) shall be as small as $\sim$1.0 $\lambda/D$
\item The optimal correction zone does not need to be circularly symmetric around the star, i.e. the correction may be optimized for one "side" of the star only (as long as the position angle can be freely chosen). However, if an IFU is implemented, circular symmetry would be desirable. A trade-off should be made between coronagraphic performance across the search field and corrected search area in order to minimize telescope time during the search phase.
\end{itemize}

Even though such a coronagraph is not implemented yet, several solutions have been proposed to achieve high contrast at a few $\lambda/D$ with ground-based telescopes: dual-zone phase mask coronagraphs \citep{Delorme2016}, vortex coronagraphs possibly combined with dedicated pupil apodizers \citep{Mawet2013,Bottom2016}, or even apodizing phase plates \citep{Otten2014}. Without entering a detailed coronagraph selection here, we note that the required pupil and focal planes are indeed available in SPHERE, which opens the opportunity to include them in the future. In terms of contrast, as we noted earlier, the performance will be limited by residual aberrations of the incoming beam (unlike space-based applications where the intrinsic ultimate coronagraphic properties more directly drive the final limitations).

\subsubsection{SPHERE fiber interface module}
\label{SectIFU}

\begin{figure}
\centering
\includegraphics[width=\columnwidth]{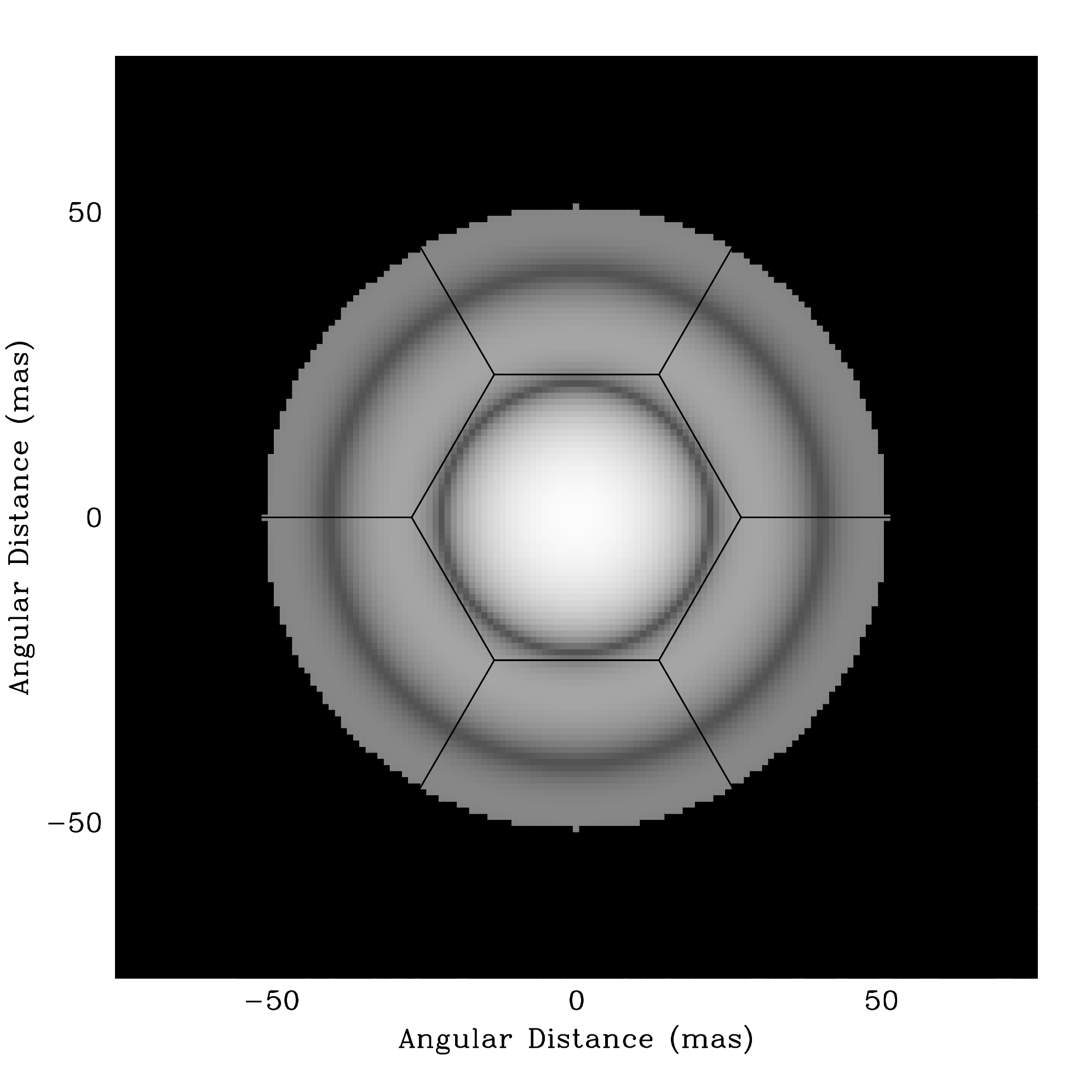}
\caption{Example of an hexagonal integral-field unit (IFU) covering the search annulus around the star. An outer field mask is included to minimize stellar contamination within IFU spaxels.}
\label{FigIFU}
\end{figure}

The present layout of the SPHERE optical bench leaves little room to directly implement a pick-up interface in the visible channel upstream of the ZIMPOL instrument. However, a number of technical solutions can be envisaged, from a pick-off mirror redirecting the light to a new module, to a full exchange scheme between ZIMPOL and the new module, where the two would be interchangeable. We leave this discussion to a more in-depth technical study, and simply assume here that the new module can be inserted at a suitable location in the visible beam.


As already discussed in Sect.~\ref{SectProxima}, the detection and characterization of Proxima b (or any other RV-detected planet) with the HCHR approach will have two distinct phases:

\begin{enumerate}
\item A search phase, where the planet location is not yet known; the fibers or IFU are positioned so as to simultaneously cover the largest possible fraction of the search annulus
\item A stare phase, where the planet location is known; one fiber or spaxel is positioned on the planet, while the other fibers/spaxels record a high-SNR star-only spectrum for reference
\end{enumerate}

In order to minimize telescope time during the search phase, it is obviously necessary to maximize the number of fibers/spaxels. However we are limited here by the accepted field of view at the entrance of ESPRESSO. We discuss this in more details in Sect.~\ref{SectESPRESSO} below, but here we already anticipate that a multiplexing of 6 can be achieved. A natural possibility is then to foresee an hexagonal IFU which would tile the search area of interest around the star. Fig.~\ref{FigIFU} shows the footprint such an IFU would have. Note that the central position (the star itself) does not need to be covered. Therefore, a total of 6 hexagonal spaxels with side length $\sim$27 mas would efficiently cover the entire search area, which is an annulus with inner and outer radii of $\sim$23 and 51 mas, respectively. The IFU could be realized using an hexagonal lenslet array that would then feed a corresponding fiber bundle. An outer field mask can be added in front of the lenslet array to minimize stellar contamination within the IFU spaxels (see Fig.~\ref{FigIFU}).

We note here that the limited multiplexing of 6 implies that one IFU spaxel covers one sixth of the search annulus, i.e. $\sim$1100 mas$^2$. In terms of SNR on the planet, this is less optimal than the 14-mas circular fiber considered in Sect.~\ref{SectAO}, which covers only 616 mas$^2$. Stellar contamination is thus increased by a factor of $\sim$1.8 in the search phase, and the $K$ factor is correspondingly reduced. In the SPHERE+ setup that would mean $K$ $\sim$ 3000.

While Proxima b can be detected in one shot with such an IFU (i.e. with a single IFU orientation), it is probably necessary to determine its position angle with more precision to optimize IFU orientation for the stare phase. It is then desirable to repeat the search observations with the IFU rotated by 30 deg with respect to the field, which would redundantly cover the whole search annulus and enable a precise location of the planet.

In the stare phase, one ideally needs to have one 14-mas aperture exactly centered on the planet to maximize SNR. Therefore, one could envisage a different field mask for this phase, with a circular aperture exactly on the planet and no obstruction at all on the other 5 spaxels to collect a high-SNR star-only spectrum.

We also note that the positioning and tracking accuracy of the IFU module in the SPHERE focal plane during an exposure shall be a small fraction of spaxel size, i.e. a few mas.

The IFU described here is specifically tailored to Proxima b. However, it actually covers the area closest to the star in an optimal way, and could be used on other interesting planets as well. For example, GJ 876 b has a slightly larger maximum elongation than Proxima b (43 mas, see Fig.~\ref{FigReflectedLight}), but a dedicated field mask with a larger outer radius could be used in this case. To further increase flexibility and field coverage, it may also be possible to design a larger fiber bundle with more spaxels, with the capability to choose which 6 spaxels to feed into the spectrograph.

\subsubsection{Injection into ESPRESSO}
\label{SectESPRESSO}

The fibers carrying the SPHERE beams will run from the UT3 Nasmyth platform A to the Coud\'e room below the telescope, where the ESPRESSO UT3 Coud\'e train optics are located. From there, two options can be envisaged to reach the spectrograph: either an all-optics solution using the existing ESPRESSO Coud\'e train, or a long-fiber solution. 

The long-fiber option is probably simpler and more efficient: the SPHERE fiber bundle would extend all the way to the spectrograph and feed it directly, bypassing the ESPRESSO Coud\'e train optics and Front-End Unit entirely. In fact, the main functionalities of the front-end are not relevant to the SPHERE+ESPRESSO combination (toggling between UTs, ADC, field/pupil stabilization). Moreover, proper injection of standard calibration sources into the different IFU spaxels requires a dedicated module that is not presently foreseen in the front-end. This module would likely be installed on the SPHERE side. In this long-fiber solution, the required fiber length would be about 100 m. At the red optical wavelengths we are interested in ($\gtrsim$ 600 nm), we expect light losses within the fiber link to be comparable, if not inferior, to the all-optics solution. For all these reasons we adopt the long-fiber option in the remainder of this paper.

A notable difference between diffraction-limited (SPHERE) and seeing-limited (ESPRESSO) observations is the geometrical etendue of the beam, which is much smaller in the diffraction-limited case. While ESPRESSO has been designed to accept two 1.0-arcsec fibers (standard high-resolution mode) or two 0.5-arcsec fibers (ultra-high resolution mode), the SPHERE fiber interface will only pick up a small number of $\sim$0.028-arcsec apertures. This represents a massive reduction in geometrical \'etendue. In principle, this would enable the use of a much smaller spectrograph (in physical size) offering the same spectral resolution as ESPRESSO. However, since ESPRESSO is already available at the telescope, its use should obviously be investigated first. The specific question that arises is then how to properly couple the SPHERE diffraction-limited beams into the large ESPRESSO fibers. We merely note here that the long fiber will ensure a high homogeneity and stability of the fiber near-field at spectrograph entrance. Dedicated simulations will show if further homogenization is necessary.

One remarkable advantage of the diffraction-limited SPHERE+ESPRESSO combination is the possibility to feed the 0.5-arcsec fibers of ESPRESSO ($R$ = 220,000) with no flux losses compared to the 1.0-arcsec fibers, thereby achieving an extremely high spectral resolution "for free". Given the narrow width of the spectral lines in the Proxima spectrum, and the similarly narrow width of planetary spectral lines and telluric lines ($\lesssim$ 3 km s$^{-1}$), an ultra-high resolution of $R$ = 220,000 is an important advantage in terms of line contrast, and thus signal, and telluric rejection. It also has the additional benefit of illuminating less pixels on the ESPRESSO detectors, thereby minimizing readout and dark current noise. Correspondingly, we envisage here to use the slow-readout detector mode of ESPRESSO with 1x2 pixel binning (i.e. binning in cross-dispersion direction). The resulting FWHM of the spectral orders will then be about 2 binned pixels. This mode will yield a readout noise of about 6 electrons per extracted pixel. Dark current noise will amount to only 2 electrons per hour per extracted pixel.

We now turn to the critical question of how many IFU spaxels could be simultaneously fed into ESPRESSO. The situation is made relatively complex because of the anamorphic pupil slicer unit (APSU) located at the entrance of the spectrograph, which accepts only two rectangular fields of view and performs pupil slicing on those. Correspondingly, only two fibers per ESPRESSO observing mode can be simultaneously fed into the spectrograph in the existing configuration. However we suggest here that a minor modification to the fiber injection interface would enable a higher multiplexing in the ultra-high resolution (UHR) mode. Basically, taking advantage of the small size of the fibers in the UHR mode, it is possible to vertically align 3 input fibers per field of view, i.e. 6 fibers in total, without changing anything to the spectrograph optics and APSU in particular. Indeed, the maximum height of the field of view is 280 $\mu$m at the entrance focal plane, while the UHR fibers are 70 $\mu$m in diameter. Aligning 3 of them vertically would then leave a $\sim$35 $\mu$m spacing between them. This would correspond to at least 2.5 pixels of vertical separation on the detector. Given that HCHR observations will always be low-SNR measurements (minimizing stellar light to reveal the planet), and all spaxels will roughly have the same flux level, we expect that cross-talk between adjacent fibers on the detector will be negligible. In the foreseen configuration, the 6 input IFU spaxels will produce 2 groups of 3 interleaved spectra on the ESPRESSO detectors, with ample spacing separating the 2 groups (corresponding to the original spacing between fibers A and B). Thus, there will still be many non-illuminated pixels available for background measurement (e.g. scattered light, bias residuals, dark current).

In conclusion, we propose to add a new mode to ESPRESSO, the UHR-IFU mode, in which a bundle of 6 UHR fibers feed the spectrograph. Those would be vertically aligned in 2 groups of 3 fibers at the entrance focal plane of the spectrograph. In the long-fiber solution discussed above, the fibers would come directly from SPHERE without going through the ESPRESSO front-end. The implied modifications to the existing ESPRESSO instrument are relatively minor; in particular, nothing has to be changed inside the spectrograph itself.

\subsubsection{System throughput}

We estimate the throughput of the SPHERE+ESPRESSO channel by following the photons from the top of the atmosphere to the ESPRESSO detector (see Table \ref{TableTrans}). We assume standard values for the atmospheric transmission and telescope throughput at a wavelength of 700 nm. The throughput of the SPHERE Common Path Infrastructure (CPI) is about 50\% \citep{Dohlen2016}. Note that we assume here that all the light between 600-780 nm goes to ESPRESSO while all the light beyond 780 nm is used for wavefront sensing, i.e. we assume an optimized beamsplitter in the SPHERE visible arm. The coronagraph is assumed to have a transmission of 70\%. In the case of the short-term SPHERE upgrade, the fiber coupling efficiency is 38\% assuming a 14-mas fiber (Sect.~\ref{SectAO}). For the more ambitious SPHERE+ that we consider here, the coupling efficiency is assumed to be 60\%.

We further assume that a long fiber bundle directly links SPHERE to ESPRESSO, bypassing the ESPRESSO Coud\'e train and Front-End Unit. The light from the fiber bundle is injected directly into the spectrograph, without any further scrambling optics. Using a typical value of 10 dB km$^{-1}$ for light attenuation within the fiber at 700 nm, and assuming a 100-m long fiber, one then obtains a throughput of 80\% for the fiber link.

\begin{table}
\caption{Transmission budget for SPHERE+ESPRESSO, assuming SPHERE+ and a long-fiber link directly to ESPRESSO.}
\label{TableTrans}
\centering
\begin{tabular}{lc}
\hline\hline
Item & Throughput \\
\hline
Atmosphere & 0.97 \\
Telescope & 0.65 \\
SPHERE CPI & 0.50 \\
Coronagraph & 0.70 \\
Fiber Coupling & 0.60 \\
Fiber Link & 0.80 \\
Spectrograph & 0.40 \\
\hline
Total & 0.042 \\
\hline
\end{tabular}
\end{table}

On the ESPRESSO side, we use a total throughput of 40\% for the spectrograph including the detector \citep{Pepe2014}. Overall, we obtain a total throughput of 4.2\% for SPHERE+ESPRESSO in the SPHERE+ case. This number would decrease to 2.7\% in the short-term upgrade case.

\section{Detecting Proxima b and its atmosphere}

\subsection{Simulations of Proxima observations}
\label{SectSimul}

\begin{figure*}
\centering
\includegraphics[width=15cm]{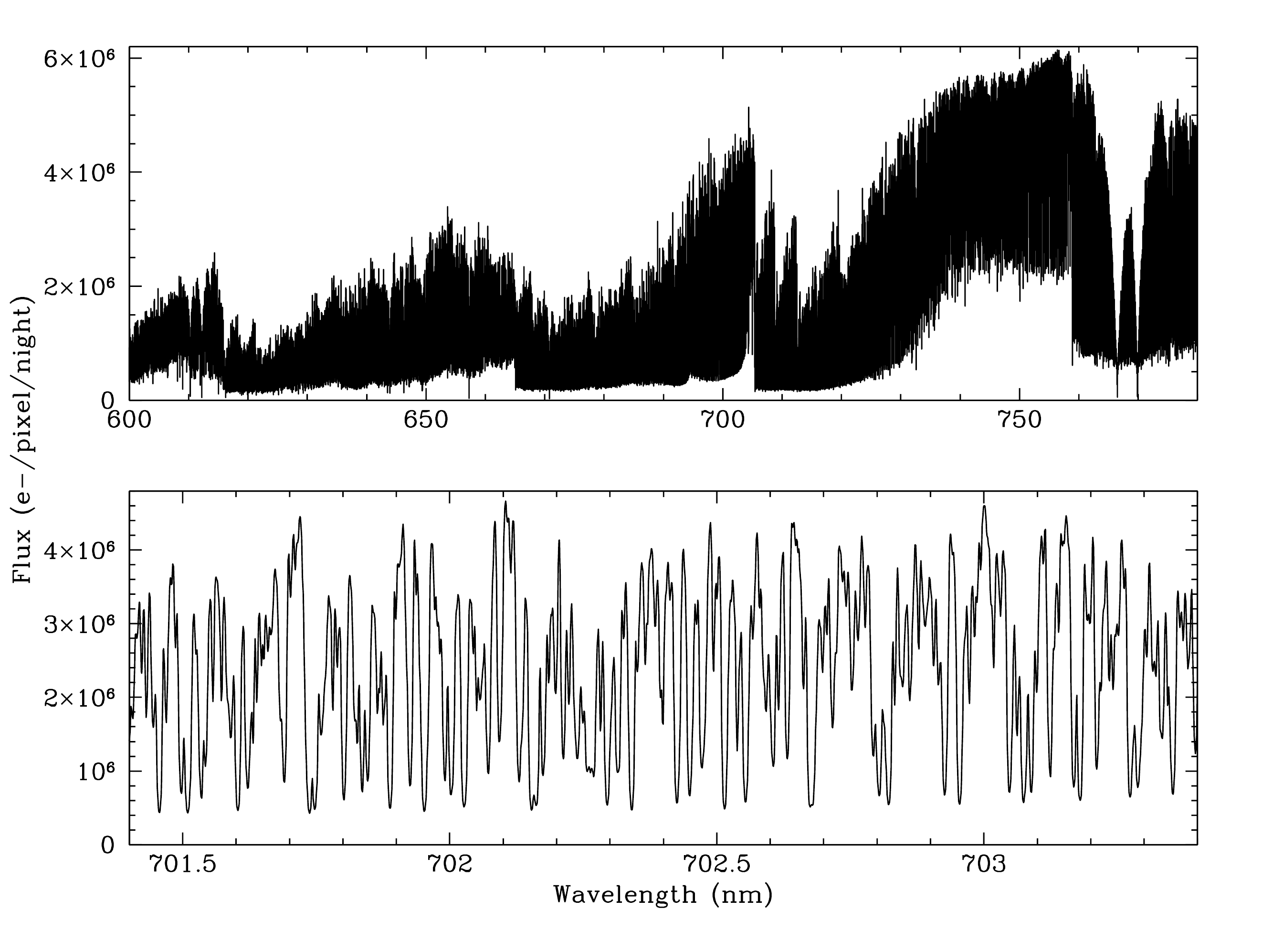}
\caption{PHOENIX high-resolution spectrum simulating an observation of Proxima with SPHERE+ESPRESSO. The total integration time is 8 hours. Top: the 600-780 nm region, which would be used for SPHERE+ESPRESSO observations. Bottom: close-up view showing the very high density of spectral lines in this star.}
\label{FigPHOENIX}
\end{figure*}

We simulate the high-resolution spectrum of Proxima using the PHOENIX stellar spectral library \citep{Husser2013}. We choose the stellar model with $T_{\mathrm{eff}}$ = 3000 K, $\log g$ = 5.50 and [Fe/H] = 0.0, because these parameters are close to the measured ones \citep{Demory2009}, and the stellar luminosity for this model, as given by PHOENIX, matches the observed luminosity of Proxima to within a few percent \citep[0.00155 $L_\odot$ or 5.93 $\times$ 10$^{30}$ erg s$^{-1}$;][]{Boyajian2012}. We convert the PHOENIX fluxes at the stellar surface into a number of photo-electrons per extracted pixel, as detected by the SPHERE+ESPRESSO instrument at the VLT (an extracted pixel in ESPRESSO corresponds to about 0.45 km s$^{-1}$ in velocity units, or 0.001 nm at 700 nm).

We do not apply any rotational broadening to the simulated spectrum because it is negligible in the case of Proxima. Indeed, considering a stellar radius of 0.14 $R_\odot$ and a stellar rotation period of 83 days \citep{Anglada2016}, one obtains an equatorial velocity of only 0.085 km s$^{-1}$.

Fig.~\ref{FigPHOENIX} shows the simulated Proxima spectrum in the 600-780 nm wavelength range. Fluxes are detected photo-electrons per extracted pixel for a total integration time of 8 hours (i.e. 1 night), assuming the SPHERE+ throughput. As can be seen, the average pseudo-continuum flux in this region is about 3.0 $\times$ 10$^6$ electrons per spectral bin. This is the flux that would be obtained with the pick-up fiber centered on the star. We can now estimate the flux on the planet spectrum. Assuming a planet-to-star contrast of 1.0 $\times$ 10$^{-7}$ (see Sect.~\ref{SectReflectedLight}), this means that the average reflected flux from Proxima b amounts to just $\sim$0.3 electron per pixel over 8 hours of observations. Expressed in magnitudes, we are attempting to detect a source that is 17.5 magnitudes fainter than its host star, i.e. at an $I$-band magnitude of 24.9. At first glance it may sound like a daunting task just because of this extreme faintness, especially when using a high-resolution spectrograph. However, we suggest an analogy here to put things into context: in the case of radial velocity observations of exoplanets, moonlight contamination sometimes affects the precise RVs of faint targets by contributing a weak reflected solar spectrum on top of the target spectrum, thereby producing a similar situation as our star+planet observation. Taking HARPS as an example, it is not uncommon to detect the cross-correlation (CCF) peak of the Moon at a continuum level of $\sim$2000 electrons per CCF velocity bin in a bright-time observation. Given that about 3600 spectral lines are used to build the CCF, the corresponding flux in the spectrum is at the level of $\sim$0.5 electron per extracted pixel (the CCF is basically a co-addition of all lines in velocity space). In such a case the CCF peak of the Moon is clearly detected despite a readout noise of 12 electrons per pixel. Therefore, the weakness of the signal from Proxima b (in absolute terms) is not in itself an obstacle. We show below that a detection in these circumstances is indeed possible with the HCHR technique using SPHERE+ESPRESSO.

\subsection{Noise estimates}

Following Eq.~\ref{EqSNR2}, we first need to assess the magnitude of all relevant noise sources to compute the achievable SNR on the planet spectrum. In Sect.~\ref{SectESPRESSO}, readout noise and dark current noise were estimated to be 6 and 2 electrons per pixel per 1-hour exposure, respectively. Sky background (and its associated noise) is usually what sets the limiting magnitude for faint object detection, so we need to carefully examine its impact. We first note the special case of our HCHR observations regarding sky background: we are collecting visible light from a tiny patch of sky (0.028 arcsec in diameter) and sending it to a high-resolution spectrograph. We proceed based on the very detailed Cerro Paranal Sky Model developed at ESO \citep{Noll2012,Jones2013}, which includes contributions from telluric emission lines, scattered moonlight, scattered star light, zodiacal light, and airglow continuum. Contamination by telluric emission lines can be neglected in high-resolution spectroscopy since the lines are intrinsically very narrow (a few km s$^{-1}$) and only affect a small fraction of the spectrum in the visible domain. To evaluate the continuum contribution we run the ESO Sky Model tool using input parameters representing realistic Proxima observations. In dark time (new Moon), typical continuum fluxes at 700 nm are 250 photons s$^{-1}$ m$^{-2}$ $\mu$m$^{-1}$ arcsec$^{-2}$, while in bright time (full Moon) those increase to about 3000 photons s$^{-1}$ m$^{-2}$ $\mu$m$^{-1}$ arcsec$^{-2}$. We convert these values into detected fluxes by SPHERE+ESPRESSO under the same conditions as the Proxima simulation above (except that we do not include throughput losses at fiber injection, since sky background is not a point source). We obtain fluxes of 0.015 and 0.18 electron per pixel over 8 hours at new Moon and full Moon, respectively. Corresponding shot noise contributions are negligible. In dark time, we thus see that sky background is significantly fainter than Proxima b itself, while in bright time the two contributions are comparable. Since in bright time the sky background is dominated by moonlight contamination, which is a reflected solar spectrum that is well-known, it is in principle possible to model and subtract this contribution accurately (the 5 IFU spaxels that are not on the planet would provide an empirical model). However it is probably necessary to require photometric conditions when the Moon is up, to avoid additional scattering by clouds and cirri. In conclusion, the sky background signal must be taken into account in bright time for the purpose of detecting Proxima b, while sky background {\it noise} is always negligible compared to detector readout and dark current noise.

The remaining noise source to be quantified is photon shot noise from the residual star light at the planet location. As shown in Eq.~\ref{EqSNR2}, this directly depends on the $K$ factor. We examine how stellar photon noise compares to other noise sources in a 1-hour exposure with SPHERE+ESPRESSO, the maximum practical duration of an exposure at the VLT. When the fiber is centered on the star, the average flux of Proxima in 1 hour is about 2.3 $\times$ 10$^5$ electrons per pixel (SPHERE+ assumed). For $K$ = 1000, the stellar flux at the planet location is reduced to 230 electrons, yielding a shot noise of 15 electrons. For $K$ = 5000, the shot noise decreases to just 7 electrons. Thus, in summary, there are significant contributions to the noise budget from stellar photon noise, detector readout noise (6 electrons), and dark current noise (2 electrons). Stellar shot noise dominates up to $K$ $\sim$ 5000, at which point detector-related noise takes over and sets the noise floor.

\subsection{Detecting the reflected spectrum}
\label{SectDetectability}

We can now estimate the amount of observing time that is required to detect Proxima b. Hereafter we will assume that we are working with SPHERE+, i.e. with increased throughput and a $K$ factor of 3000 in search mode.

In reflected light, the planet is detected if we manage to pull out the cross-correlation (CCF) signal of the reflected spectrum using the direct stellar spectrum as a template. We compute the achieved SNR on the planet CCF by performing a realistic cross-correlation simulation and calibrating Eq.~\ref{EqSNR2} based on the simulation results. We start from the PHOENIX spectrum described above and simulate the reflected spectrum from Proxima b by applying a Doppler shift and a number of possible contrast values. Regarding the Doppler shift, we note that the orbital velocity of the planet is known from the RV orbit and is equal to 47 km s$^{-1}$. At quadrature, the planet cross-correlation signal will thus be clearly separated from the direct stellar spectrum in velocity space for all but the smallest orbital inclination angles. For the present simulation we assume an observed Doppler shift of 43 km s$^{-1}$, corresponding to a slightly inclined orbit. Planet-to-star flux ratios at quadrature are taken from \citet{Turbet2016} and range from 4 $\times$ 10$^{-8}$ (dry planet with CO$_2$-rich atmosphere) to 2 $\times$ 10$^{-7}$ (Venus-like atmosphere with aerosols). Earth-like atmospheres show contrasts between 1.0 and 1.4 $\times$ 10$^{-7}$. We then add the direct stellar spectrum according to the adopted $K$ factor of 3000. We also include telluric absorption (see Sect.~\ref{SectBio}). We generate Gaussian noise on each pixel according to the total noise derived from the quadratic sum of photon noise, readout noise, and dark current noise. We subsequently cross-correlate the noisy planetary spectrum with the original stellar template, assuming that the direct stellar spectrum can be perfectly removed. The latter assumption is justified by the availability of 5 simultaneously-obtained star-only spectra that provide a high-SNR reference. The SNR on the planetary CCF peak is estimated by dividing the peak amplitude by the scatter in the continuum around it. Finally, we generate a large number of simulations from which we calibrate the constant factors in Eq.~\ref{EqSNR2} (i.e. leaving only the planet contrast and exposure time as variables).


\begin{figure}
\centering
\includegraphics[width=\columnwidth]{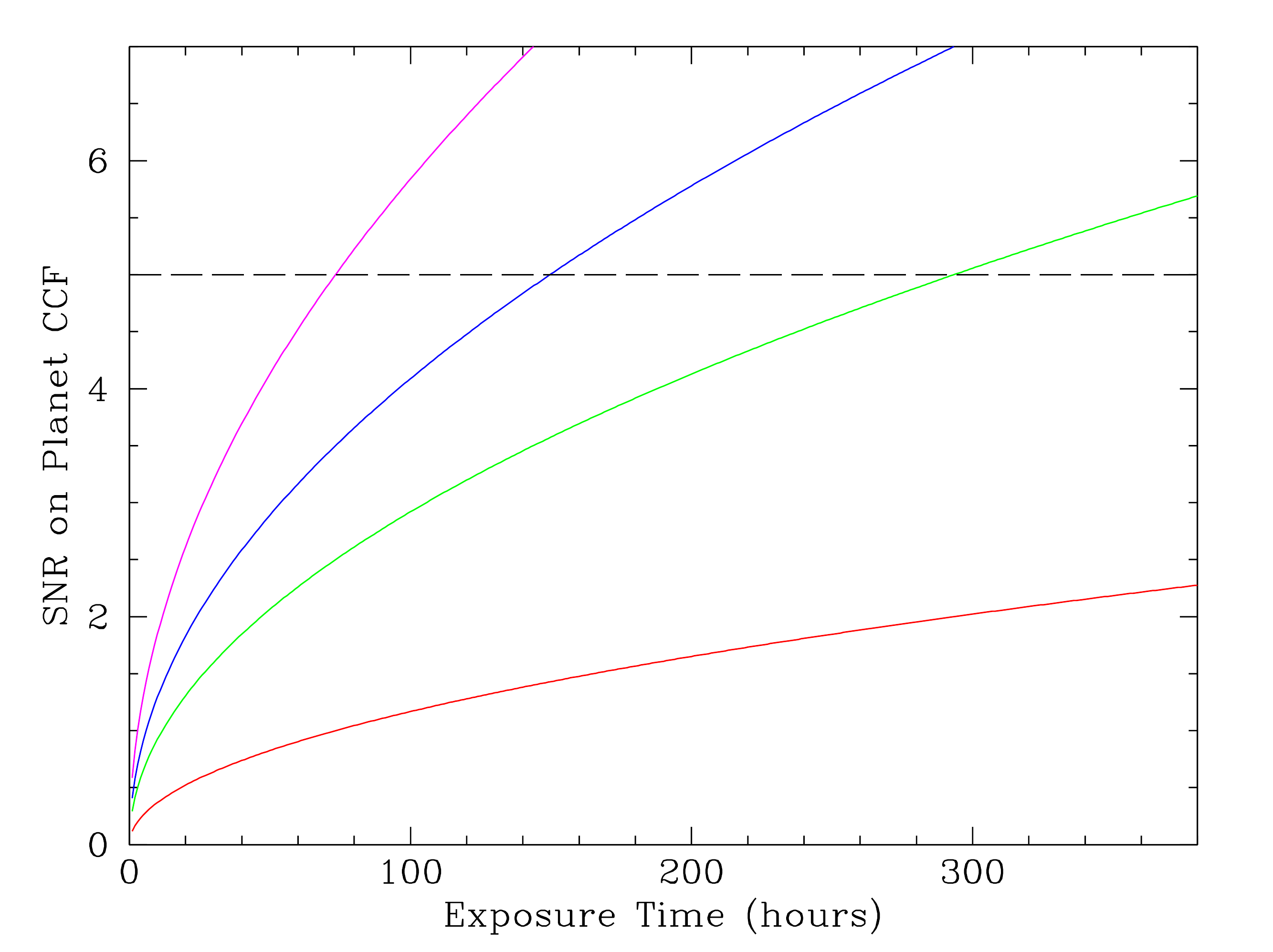}
\caption{Detectability of the reflected spectrum from Proxima b with SPHERE+ESPRESSO. The achievable SNR on the planetary cross-correlation function (CCF) is plotted as a function of exposure time. Four different planet-to-star contrasts from \citet{Turbet2016} are considered. From bottom to top: 4 $\times$ 10$^{-8}$ (dry planet with CO$_2$ atmosphere), 1 $\times$ 10$^{-7}$ (synchronous Earth-like planet), 1.4 $\times$ 10$^{-7}$ (Earth-like planet in 3:2 resonance), and 2 $\times$ 10$^{-7}$ (Venus-like atmosphere with aerosols). The horizontal dashed line indicates SNR = 5.}
\label{FigSNR}
\end{figure}

Fig.~\ref{FigSNR} shows the results of our SNR calculations. The achieved SNR on the planet CCF is plotted as a function of exposure time for the four contrast values described above. We see from this figure that Proxima b can be detected at 5-$\sigma$ in about 19 and 37 nights of VLT time for the two Earth-like cases (3:2 resonance and synchronous rotation, respectively) with SPHERE+ESPRESSO. A Venus-like atmosphere can be detected at 5-$\sigma$ in only 9 nights. This is assuming a SPHERE+ performance level. We note here that the SNR on the planet spectrum varies with the square root of the instrument throughput and $K$ factor (Eq.~\ref{EqSNR3}). Thus, detecting Proxima b with the reduced performances of the short-term SPHERE upgrade ($K$ $\sim$ 500, lower throughput) would be extremely demanding in terms of observing time, if not impossible.

We conclude that SPHERE+ESPRESSO with the SPHERE+ upgrade will be able to detect Proxima b in a reasonable amount of time if it has an Earth-like atmosphere or a surface/atmosphere with similar or higher reflectivity. Note that we are considering here time-averaged reflectivity values, as the amount of observing time needed for a detection will smear out short-term temporal variations (i.e. weather).

\subsection{Observing and finding Proxima b}
\label{SectFinding}

As described in Sect.~\ref{SectProxima}, the RV orbital solution provides us with the timing of both quadratures within each 11.2-d orbit of Proxima b. Those are the only orbital phases where a detection with SPHERE+ESPRESSO can be attempted. In the worst case of an edge-on orbit ($i$ = 90 deg), the planet would spend 1.6 day (39 hours) at a separation larger than 90\% of the maximum elongation (i.e. 33 mas) at each quadrature. With 5-6 quadratures occurring each month, there are a total of about 8.6 favourable observing nights per month. Proxima, at a southern declination of -63 deg, is well observable from Paranal. The star culminates in May, when it is observable for 9.5 hours per night at an airmass below 2.0. Moreover, it is accessible for at least 6 hours per night from early March to early July, i.e. over 4 months. In total this amounts to about 34 optimal nights per year, during which the star is well observable and the planet is at maximum elongation.

As a consequence, considering the required number of nights from the previous section, the planet can be detected in about one observing season if it has the reflectivity of an Earth-like atmosphere (or higher).

In passing, we note that the 1-hour integrations will not significantly broaden the spectral features in the reflected spectrum. Indeed, we calculate a maximal radial acceleration of $\sim$0.5 km~s$^{-1}$~hour$^{-1}$ for the planet in the 39-hour window we consider around each quadrature. Given the typical line width of $\sim$3 km~s$^{-1}$ in the stellar spectrum, this leads to negligible smearing.
 
Practically speaking, finding Proxima b means determining the position angle of the planet at quadrature. Fortunately, thanks to the IFU capability described in Sect.~\ref{SectIFU}, SPHERE+ESPRESSO will be able to cover the entire search area around the star in a single shot. However, the planet PSF may extend over two spaxels, which is not optimal in terms of SNR (more stellar photon noise). Therefore, it may be desirable to more precisely measure the planet position by repeating the observations with the IFU rotated by 30 deg. This would redundantly cover the full search area and significantly improve the precision on the position angle, so that further observations can be carried out with the IFU optimally oriented (and with an optimal aperture size).

\subsection{Mass and albedo measurement}

Important information on the planet can already be obtained using the "discovery" data. First, as soon as the planet reflected light is detected through cross-correlation with the stellar template, the planetary orbit will be fully defined and the planet true mass will be known. On the one hand, the longitude of the ascending node is directly measured from the detected planet position on the IFU. On the other hand, the orbital inclination and mass are obtained from the observed planet radial velocity. Indeed, at quadrature, the RV semi-amplitude of the planet is directly measured which, combined with the known RV semi-amplitude of the star, gives the planet-to-star mass ratio and thus the true planet mass and orbital inclination.

Eq.~\ref{EqReflectedLight2} shows that the measured reflected-light signal constrains other fundamental properties of the planet, namely its radius, albedo, and phase function.  However, these three quantities can only be measured together and, in particular, the radius is not known a priori since the planet is most likely not transiting \citep{Kipping2016}. Nevertheless, the mass-radius relation of known exoplanets suggests that most $\sim$1.5-$M_\oplus$ objects on short-period orbits have an internal structure similar to Earth \citep[e.g.][]{Rogers2015,Dressing2015}. Thus, given the known mass and assumed rock/iron composition, the radius of Proxima b can be estimated and geometric albedo can then be disentangled from it. The inability to measure radius directly is certainly a major limitation that affects all non-transiting planets. We simply note here that the upcoming transit-search missions TESS, CHEOPS and PLATO will bring our understanding of the mass-radius relation to a level that may allow us to indirectly constrain the radius of a planet like Proxima b.

In practice, we will be able to estimate the albedo of Proxima b from the SPHERE+ESPRESSO data by using the measured contrast of the planetary cross-correlation signal with respect to the direct stellar CCF. This contrast must then be corrected using the actual $K$ factor to recover the true planet-to-star flux ratio. This step is inherent to the HCHR technique: spatially resolving the planet from the star implies the loss of the true flux ratio between planet and star. In other words, the $K$ factor is an additional unknown that one has to calibrate if one wants to access the planetary properties (albedo and phase function) in a quantitative way. We believe it will be possible to measure the $K$ factor independently by monitoring the total flux incoming on the coronagraph in real time and calibrating the coronagraph and fiber coupling efficiencies.

Finally, we note that the data will first yield a broadband albedo that is averaged over the 600-780 nm wavelength range of the observations. Once sufficient SNR is accumulated, the range can be split into smaller chunks and finer details of the albedo spectrum can be accessed.

\subsection{Search for biosignatures in the albedo spectrum}
\label{SectBio}

\begin{figure*}
\centering
\includegraphics[width=14cm]{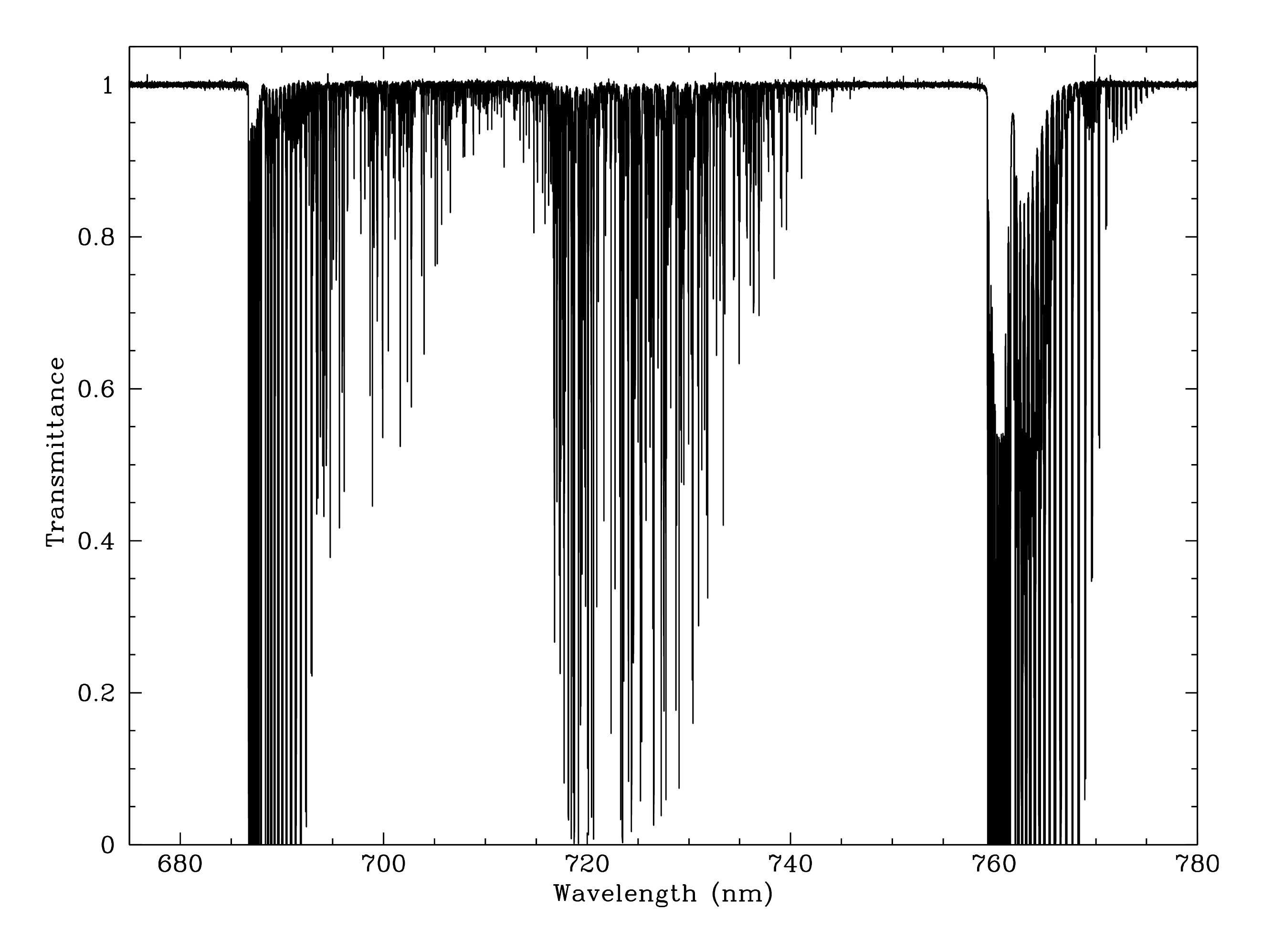}
\caption{High-resolution telluric absorption spectrum in the 680-780 nm wavelength region \citep[from][]{Hinkle2003}. Three prominent molecular bands can be seen: the oxygen B- and A-bands at 686 and 760 nm, as well as water vapour at 717 nm.}
\label{FigTelluric}
\end{figure*}

We now turn to the feasibility of detecting atmospheric spectral features on Proxima b. Starting from Eq.~\ref{EqReflectedLight2} we first write an explicit expression for the {\it observed} spectrum at the planet location:

\begin{multline}
F_{obs}(\lambda, v_s, v_p, \alpha) = T(\lambda) \frac{F_s(\lambda, v_s)}{K} 
\\ + T(\lambda) \, F_s(\lambda, v_p) \, A_g(\lambda, v_p) \, g(\alpha) \, \left(\frac{R_p}{a}\right)^2 ,
\label{EqReflectedLight3}
\end{multline}

where $T(\lambda)$ is the telluric absorption at the time of the observation, $v_s$ is the apparent radial velocity of the star, and $v_p$ is the apparent radial velocity of the planet (all three in the rest frame of the observer). Similarly, we can write an expression for the star-only reference spectrum collected by the 5 spaxels that are not on the planet:

\begin{equation}
F_{ref}(\lambda, v_s) = T(\lambda) \frac{F_s(\lambda, v_s)}{K_i} ,
\label{EqReflectedLight4}
\end{equation}

where $K_i$ simply denotes the fact that each spaxel will have a slightly different stellar light rejection factor. Critically, these 5 star-only spectra are acquired {\it simultaneously} with the star+planet spectrum, so that we can assume that the telluric transmission is the same for all spaxels. Moreover, also the intrinsic stellar spectrum is acquired simultaneously, so that any variability caused by stellar activity will be captured by all spaxels.

Detecting atmospheric features from the planet essentially means extracting the high-resolution albedo spectrum $A_g(\lambda, v_p)$ from the observed data described by Eq.~\ref{EqReflectedLight3}. In the following we attempt to do this in three steps:

\begin{enumerate}
\item Simulate a realistic observation as before, now including molecular absorption in the albedo spectrum
\item Fit the data using appropriate models for the stellar and telluric spectra, assuming a flat albedo spectrum
\item Study the residuals to the best-fit model in order to probe excess absorption in the molecular bands of interest
\end{enumerate}

A planetary spectral feature is considered detected if significant absorption is measured in the specific spectral bins where it is expected to be found, but not in adjacent bins that are slightly offset in radial velocity. This makes use of the unique high-resolution fingerprint that a particular molecule imprints on the spectrum. We choose this approach here for its generality and ease of implementation. A proper modeling would include a series of theoretical planet spectra to be included into the full model and fitted to the data; however creating an extensive library of plausible spectra for Proxima b is beyond the scope of this paper. Instead we adopt an empirical approach based on the residuals to a flat albedo spectrum, verifying that the fitted model parameters are not strongly affected by the use of such a simplified model. The general idea here is that, at high spectral resolution, the various components (stellar, telluric, planetary) exhibit spectral signatures so different from each other that they are practically orthogonal, i.e. model parameters describing different components are almost uncorrelated.

In the following we will be using the star-only spectra (Eq.~\ref{EqReflectedLight4}) as models for the stellar and telluric components. Not only are they recorded simultaneously with the planet spectrum, but they are also at significantly higher SNR, since 5 spaxels are available and can be co-added. We therefore have an almost ideal, empirical model, the use of which adds a negligible amount of noise to the noise budget. For the purpose of fitting a model to the star+planet observation, Eqs.~\ref{EqReflectedLight3} and \ref{EqReflectedLight4} can thus be combined and rewritten as follows:

\begin{equation}
F_{obs}(\lambda) = \gamma_1 \, F_{ref}(\lambda) + \gamma_2 \, T(\lambda) \, F_s(\lambda, \gamma_3) ,
\label{EqModel}
\end{equation}

where $\gamma_1$, $\gamma_2$ and $\gamma_3$ are the three free parameters of the model, corresponding to the effective $K$ factor, the mean planet-to-star flux ratio, and the radial velocity of the planet, respectively. Since the reflected stellar spectrum is Doppler-shifted with respect to direct stellar light, it is necessary to obtain a separate model for the stellar spectrum and the telluric absorption. The best procedure to do this has to be studied; we merely note here that Proxima spectra obtained at various epochs during the year can be used to "fill the gaps" masked by telluric absorption, taking advantage of the variable barycentric correction. A clean stellar spectrum can then be reconstructed, from which the telluric contribution can be disentangled.

\subsubsection{Oxygen}

We now proceed with an attempt at detecting molecular oxygen on Proxima b, assuming the planet is an Earth analog in terms of atmospheric composition. For the simulation we need a realistic, high-resolution telluric (and oxygen) spectrum. We choose the FTS scans obtained at the McMath Fourier Transform Spectrometer at Kitt Peak \citep{Hinkle2003}. Fig.~\ref{FigTelluric} shows telluric absorption as seen after a single, vertical pass through the atmosphere. Three prominent molecular bands can be seen in the 680-780 nm region: the oxygen B- and A-bands at 686 and 760 nm, as well as water vapour at 717 nm. An additional oxygen band ($\gamma$-band, not shown in Fig.~\ref{FigTelluric}), is present at 627 nm, although weaker. We use this spectrum as telluric model, scaling it to an airmass of 1.4 to mimic a real Proxima observation from Paranal.

Hereafter we simulate the albedo spectrum $A_g(\lambda)$ of Proxima b by simply scaling this same telluric spectrum to an effective aimass of 2.83. The chosen airmass matches the column density seen by a light ray going through the planetary atmosphere in double pass at an incidence angle of 45 deg (quadrature geometry). The full reflected spectrum is then constructed by multiplying with an assumed broadband planet-to-star contrast of 1.4 $\times$ 10$^{-7}$, and the stellar spectrum. Finally, the relevant Doppler shift is applied to the reflected spectrum.

Obviously, a major obstacle to O$_2$ detection in an exoplanet is the presence of the very same spectral lines in Earth's atmosphere. Fortunately, using high-resolution spectroscopy makes it possible to disentangle the telluric and planetary line systems thanks to their different Doppler shifts. We thus need to first examine when the planetary spectral lines are well separated from the telluric ones. As an example, \citet{Rodler2014} derive the optimal relative RVs between the planetary and telluric O$_2$ A-bands (their Fig.~4), showing that relative RVs of $\pm$ (15-30) km s$^{-1}$ are particularly suitable. In general, the relative RV is given by:

\begin{equation}
\Delta RV = RV_{sys} + RV_{orb} - RV_{bary},
\label{EqRelRV}
\end{equation}

where $RV_{sys}$ = -21.7 km s$^{-1}$ is the systemic RV of the Proxima system, $RV_{orb}$ is the projected orbital velocity of the planet with respect to the Proxima system barycenter, and $RV_{bary}$ is the projected velocity of the observer with respect to the Solar System barycenter. Since observations are obtained at quadrature, $RV_{orb}$ is equal to $\pm$ 47 km s$^{-1}$ for $i$ = 90 deg. Here we adopt a slightly smaller $RV_{orb}$ = $\pm$ 43 km s$^{-1}$ assuming the orbit is not perfectly edge-on. As discussed in Sect.~\ref{SectFinding}, observations will be carried out in the March-July period, when $RV_{bary}$ is between +20 and -15 km s$^{-1}$ (going through 0 km s$^{-1}$ in mid-May). In summary we have:

\begin{equation}
\Delta RV = [-1, +34] \; \mathrm{km \,s^{-1} \; (receding \;quadrature)}
\label{EqRelRV1}
\end{equation}

\begin{equation}
\Delta RV = [-83, -48] \; \mathrm{km \,s^{-1} \; (approaching \;quadrature)}
\label{EqRelRV2}
\end{equation}

We find that relative RVs are too close to zero at receding quadrature at the beginning of the observing season, but the situation becomes ideal from early May onwards. At approaching quadrature the situation is inverted. Line blending increases significantly around -50 km s$^{-1}$, which corresponds to the typical spacing of the A-band lines \citep{Rodler2014}. Thus, good relative RVs prevail until early June only. In summary, observations should target the planet at approaching quadrature in March-May, and receding quadrature in May-July. The number of optimal observing nights per year is correspondingly reduced from 34 to about 23 nights, although note that we required at least 6 hours of observability per night, which can be relaxed to some extent. In the following simulation we assume a relative RV of -80 km s$^{-1}$, the planet being at approaching quadrature.

We create a set of observed star+planet spectra following Eq.~\ref{EqReflectedLight3}, adding photon, readout and dark current noise as appropriate. Then, we fit the simple, flat-albedo model given in Eq.~\ref{EqModel} to the data via $\chi^2$-minimization. Finally, we search for excess absorption by O$_2$ in the residuals to the best-fit model. To do so we proceed as follows:

\begin{enumerate}
\item In the telluric template, we identify all pixels that fall within one of the three oxygen bands (627, 686 and 760 nm), and which show a transmittance of less than 90\%. This array of wavelengths represents the high-resolution O$_2$ spectral fingerprint, where the O$_2$ signal is concentrated. It contains 6.2\% of the 600-780 nm spectral range.
\item We shift the spectral fingerprint according to the {\it known} radial velocity of the planet
\item We sum the residuals to the best-fit model that fall within the shifted spectral fingerprint. This yields the integrated O$_2$ absorption from the planet.
\item We repeat this operation using a number of arbitrarily-shifted fingerprints to obtain control fluxes. 
\end{enumerate}

\begin{figure*}
\centering
\includegraphics[width=18cm]{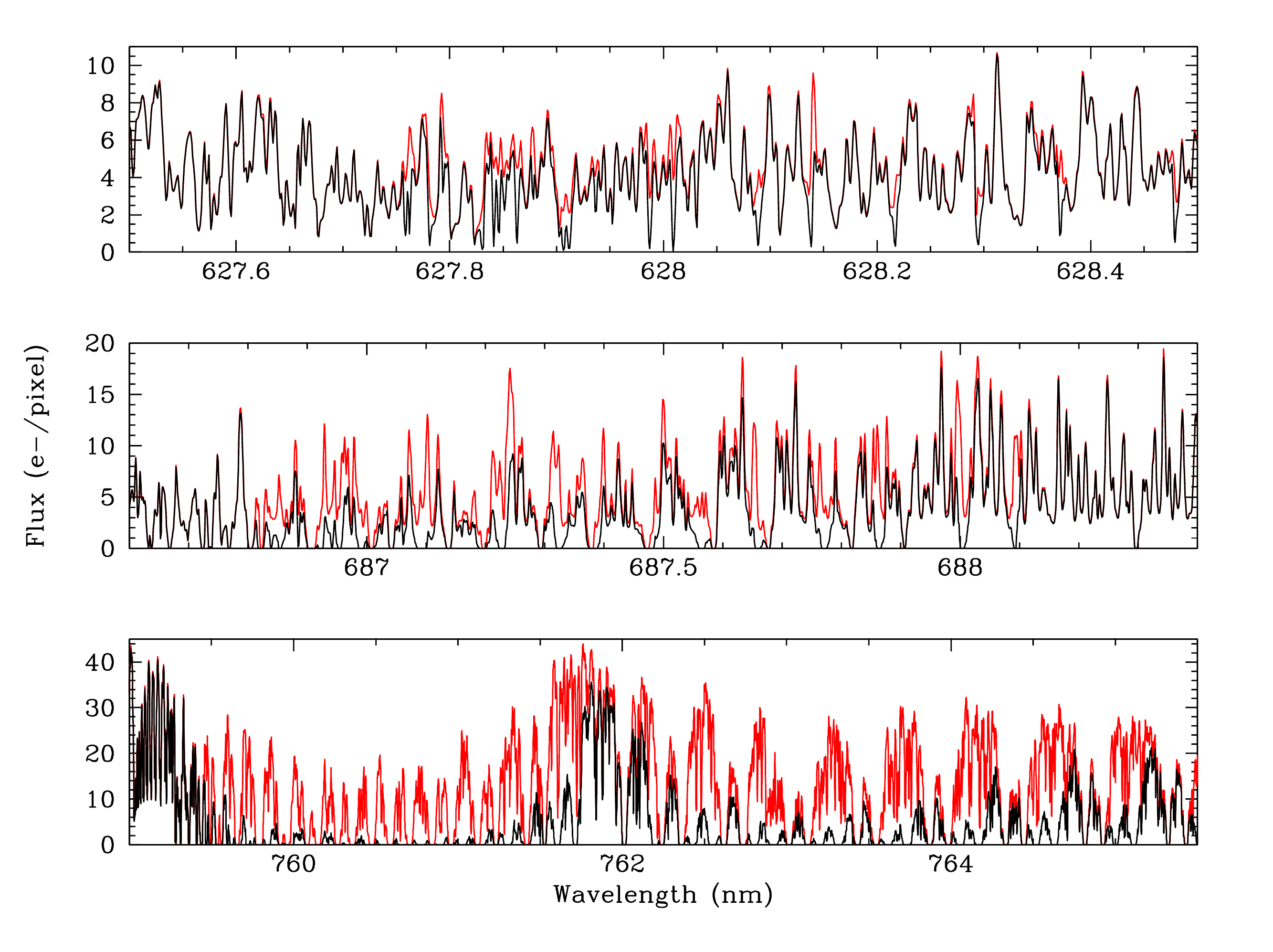}
\caption{Simulated spectrum of Proxima b in the O$_2$ bands, assuming 60 nights of observations with SPHERE+ESPRESSO. The flat-albedo model comprising only the reflected stellar component and telluric absorption is shown in red, while the full model including O$_2$ absorption from the planet is shown in black. The top, middle, and bottom panels show portions of the O$_2$ $\gamma$-, B-, and A-bands, respectively. Note that the direct stellar component, which dominates the observed spectrum, has been subtracted. The plotted model spectra are noise-free. Actual noise levels in this simulation are at the level of 200 e-/pixel (for $K$ = 5000).}
\label{FigO2}
\end{figure*}

The difference between the flat-albedo and O$_2$ models is illustrated in Fig.~\ref{FigO2} for the three available bands of O$_2$.

\begin{table}
\caption{Results of the O$_2$ detection experiment}
\label{TableO2}
\centering
\begin{tabular}{c c c c}
\hline\hline
Parameter & Injected & Mean fitted & Dispersion \\
 & value & value & \\
\hline
$K$ & 5000.00 & 4999.97 & 0.22 \\
$C$ & 1.400 $\times$ 10$^{-7}$ & 1.332 $\times$ 10$^{-7}$ & 0.088 $\times$ 10$^{-7}$ \\
$RV_{\mathrm{orb}}$ (km s$^{-1}$) & 43.48 & 43.47 & 0.15 \\
\hline
$A_{\mathrm{O2}}$ (e-) & -7.0 $\times$ 10$^{4}$ & -6.4 $\times$ 10$^{4}$ & 2.1 $\times$ 10$^{4}$ \\
$A_{\mathrm{control}}$ (e-) & 0 & 1.6 $\times$ 10$^{4}$ & 2.3 $\times$ 10$^{4}$ \\
\hline
\end{tabular}
\tablefoot{The assumed total exposure time is 480 hours (60 nights). $A_{\mathrm{O2}}$ is the planetary absorption signal by O$_2$, while $A_{\mathrm{control}}$ refers to a control window with the same fingerprint as O$_2$, but shifted by 15 nm in wavelength.}
\end{table}

The results of the whole experiment are reported in Table~\ref{TableO2}. The total exposure time in this example is 60 nights, and the $K$ factor is 5000 (i.e. assuming SPHERE+ performances with an optimal aperture centered on the planet). We report the average parameter values and dispersion from 100 different realizations of the noise. As can be seen, absorption by O$_2$ in the exoplanet atmosphere is detected at about 3.6 $\sigma$. The fitted parameters $K$, $C$, and $RV_{orb}$ show interesting patterns: while $K$ and $RV_{orb}$ are recovered at their nominal value, $C$ shows a systematic bias of $\sim$5\% towards lower values. Similarly, the total O$_2$ absorption $A_{\mathrm{O2}}$ measured on the best-fit residuals is too low by $\sim$9\% compared to the injected value. This can be understood as follows: since the fitted, flat-albedo model is incomplete, the fitting process tries to compensate the unmodelled planetary absorption by decreasing the global planet-to-star contrast $C$. The level of the direct stellar spectrum $K$ and the planetary Doppler shift $RV_{\mathrm{orb}}$ are sufficiently orthogonal to $C$ to be essentially unaffected. This demonstrates that, in the case of O$_2$ at least, we can explore the albedo spectrum of the planet by using the best-fit residuals to a flat-albedo model. Obviously, an in-depth study would include a fit to a library of planetary atmospheric models together with proper model selection, but we leave this to future work. An additional assessment of the reality of the O$_2$ detection can be obtained by probing a number of spectral control windows with the same fingerprint as O$_2$ but arbitrarily shifted in wavelength. In Table~\ref{TableO2} we show one such control window that is shifted by 15 nm compared to the planetary O$_2$ lines. As expected, the measured excess absorption in this case is compatible with zero.

We conclude that about 60 nights of VLT time at $K$ = 5000 yield a 3.6-$\sigma$ detection of oxygen, a biosignature, on Proxima b, assuming an Earth-like atmosphere and a planet-to-star contrast of 1.4 $\times$ 10$^{-7}$. Considering optimum observability conditions discussed above, these observations would take about 3 years to be completed.

\subsubsection{Water vapour}

The SPHERE+ESPRESSO wavelength range also includes absorption bands from water vapour: a weak band at 646 nm and a stronger one at 717 nm (see Fig.~\ref{FigTelluric}). We carry out the same simulation as for O$_2$ to evaluate the detectability of water vapour on Proxima b. We again use as input the McMath FTS transmission spectrum. A specificity of water vapour in Earth's atmosphere is its variability. We have not estimated the water vapour column density in the FTS spectrum, but it is most likely low to moderate by Earth standards since the spectra were obtained at Kitt Peak, a good astronomical site at an altitude of 2100 m.

We simulate H$_2$O absorption on Proxima b by scaling as before the telluric transmission to an equivalent airmass of 2.83 and applying the appropriate Doppler shift. We then retrieve the total H$_2$O absorption by summing the residuals to the best-fit flat-albedo model using the high-resolution spectral fingerprint of H$_2$O. In this case, we obtain a SNR of about 2.0 on the H$_2$O signal in 60 nights of observing time at $K$ = 5000. However, increasing the equivalent airmass (and thus water column density) by a factor of 3 would increase the SNR to about 3.3. We thus conclude that water vapour can be detected simultaneously with O$_2$ in the Proxima b spectrum provided the water column density reaches levels that are common in moderately-wet regions of Earth.

\subsubsection{Methane}

Besides O$_2$ and H$_2$O, methane (CH$_4$) is another important biosignature which also exhibits absorption features within the 600-780 nm range of SPHERE+ESPRESSO. However, to the best of our knowledge, no high-resolution line list of methane is available in this wavelength regime. We thus are not able to perform a similar simulation as for oxygen and water vapour, and limit ourselves to a qualitative discussion.

The best data available for absorption by cold methane in the visible are laboratory data obtained in the 1970s \citep[e.g.][]{Giver1978}, and observed spectra of the giant planets in the Solar System by \citet{Karkoschka1998}. These works focus on the low-resolution shape of the methane absorption bands. In Fig.~4 of \citet{Karkoschka1998} one can see a deep methane feature between 715 and 740 nm in the albedo spectra of Jupiter and Saturn. The strength of the feature appears comparable to the 717-nm H$_2$O band in the telluric spectrum. Although we are not able here to properly scale this methane band to the physical conditions of Proxima b, it is plausible that this feature is detectable with SPHERE+ESPRESSO if the planet has a methane-rich atmosphere. We stress here the importance of obtaining a high-resolution line list for this methane band to properly assess its detectability and disentangle it from the partially-overlapping H$_2$O feature at 717 nm.

\section{Conclusion}

We summarize here our main findings:

\begin{itemize}

\item We have studied the possibility to apply the HCHR technique to the temperate exoplanet Proxima b by coupling the SPHERE and ESPRESSO instruments at ESO VLT.

\item At maximum elongation, Proxima b is located 37 mas away from the star, corresponding to 2.0 $\lambda$/D at 750 nm. The expected planet-to-star contrast varies between about 4 $\times$ 10$^{-8}$ and 2 $\times$ 10$^{-7}$ at quadrature \citep{Turbet2016}. Earth-like atmospheres are predicted in the range 1.0-1.4 $\times$ 10$^{-7}$.

\item We have shown at a conceptual level that a SPHERE-ESPRESSO coupling is indeed possible. In particular, an IFU with a multiplexing of 6 is able to cover the entire search area in the SPHERE focal plane and can be fed into ESPRESSO without modifications to the spectrograph itself (only relatively minor changes at the fiber injection stage are needed).

\item The fiber coupling efficiency and $K$ factor are the critical variables to be maximized for the SPHERE+ESPRESSO combination to reach its full potential. Those are mainly controlled by the level of residual aberrations in the AO-corrected beam. Improvement of the AO correction and optimized coronagraphic solutions are needed to achieve a level of performance that enables the detection of Proxima b. This should be seen as a mid-term perspective (several years), as significant development is likely necessary. We designate this upgrade as a 2nd-generation SPHERE, or SPHERE+.

\item The $R$ = 220,000 mode of ESPRESSO can be used, coupled to slow CCD readout and 1x2 pixel binning. This offers several advantages: higher contrast in the planetary spectral features, better separation of the telluric and planetary features, less readout noise, and less dark current noise.

\item Readout noise becomes a significant contributor to the noise budget at $K$ $\sim$ 3000-5000, given an exposure time of one hour. The feasibility of taking longer exposures should be investigated to further decrease the noise level.

\item Based on a sophisticated sky model, we show that sky background will not impact the detectability of Proxima b, except in bright time where the scattered moonlight contribution may need to be taken into account.

\item We find that the reflected spectrum from Proxima b can be detected at the 5-$\sigma$ level in 20-40 nights of telescope time for a contrast enhancement factor $K$ = 3000 (SPHERE+) and a planet-to-star flux ratio of 1.0-1.4 $\times$ 10$^{-7}$ (Earth-like atmospheres). This includes a measurement of the planet true mass (as opposed to minimum mass) and orbital inclination, and the measurement of its broadband albedo.

\item We find that O$_2$ can be detected at the 3.6-$\sigma$ level in about 60 nights of observing time at $K$ = 5000, for a planet-to-star contrast of 1.4 $\times$ 10$^{-7}$. Those nights would need to be spread over 3 years to guarantee optimal observability conditions of the planet and sufficient separation between telluric and planetary O$_2$ lines.

\item We also show that H$_2$O can be probed in a similar amount of telescope time provided the H$_2$O column density is similar to wet regions of Earth.

\item Finally, it is likely that CH$_4$ is detectable as well if its column density is similar to or larger than the one seen in Jupiter and Saturn, although we could not address this point quantitatively.

\end{itemize}

In conclusion, while we do not underestimate the technical challenges of our proposed approach, we do believe that SPHERE+ESPRESSO is competitive for becoming the first instrument to characterize a habitable planet. At this point, this path appears to be a faster-track one compared to waiting for the availability of an equivalent instrument on e.g. the E-ELT. Indeed, SPHERE is an XAO system that works very well already today, including in $R$- and $I$-band. Further improvements and innovative ideas can certainly be pursued in the next years based on accumulated experience with the actual system. While ELTs will see their first light in the mid-2020s, reaching adequate AO performances in e.g. $J$-band to enable efficient HCHR observations will be a significant challenge, as well as building an appropriate high-resolution spectrograph equipped with an AO-fed IFU. Overall, we believe that the SPHERE+ESPRESSO opportunity can be seen as a useful pathfinder for similar instruments on ELTs, in particular for the E-ELT HIRES spectrograph.


Finally we note that, besides Proxima b, a number of exoplanets of various kinds could be probed with SPHERE+ESPRESSO. In particular, the gas giant GJ 876 b has an expected contrast $\sim$10 times higher than Proxima b (see Fig.~\ref{FigReflectedLight}) and would represent a relatively easy target for an early demonstration of the HCHR approach. Additional targets will be discovered around the nearest stars in the near future by ongoing and upcoming radial velocity surveys, in particular by ESPRESSO itself. SPHERE+ESPRESSO is thus by no means a single-target facility, but will enable reflected-light observations of several cool to warm exoplanets covering a broad range of masses. This is highly complementary to other planet characterization techniques such as transit spectroscopy.

\begin{acknowledgements}
This work bas been carried out in the framework of the National Centre for Competence in Research "PlanetS" supported by the Swiss National Science Foundation (SNSF). CL, FP, FW, NAD, AC, UC, DE, DS and SU acknowledge the financial support of the SNSF. PF and NCS acknowledge support by Funda\c{c}\~ao para a Ci\^encia e a Tecnologia (FCT) through Investigador FCT contracts of reference IF/01037/2013 and IF/00169/2012, respectively, and POPH/FSE (EC) by FEDER funding through the program ``Programa Operacional de Factores de Competitividade - COMPETE''. PF and NCS also acknowledge the support by FCT through project PTDC/FIS-AST/1526/2014, through national funds and by FEDER through COMPETE2020 (ref. POCI-01-0145-FEDER-016886), as well as through grant UID/FIS/04434/2013 (POCI-01-0145-FEDER-007672). PF further acknowledges support from Funda\c{c}\~ao para a Ci\^encia e a Tecnologia (FCT) in the form of an exploratory project of reference IF/01037/2013CP1191/CT0001.
\end{acknowledgements}

\bibliographystyle{aa}
\bibliography{proxima}

\begin{thebibliography}{42}
\expandafter\ifx\csname natexlab\endcsname\relax\def\natexlab#1{#1}\fi

\bibitem[{{Anglada-Escud\'e} {et~al.}(2016){Anglada-Escud\'e}, {Amado},
  {Barnes}, {Berdi\~nas}, {Butler}, \& {Coleman}}]{Anglada2016}
{Anglada-Escud\'e}, G., {Amado}, P.~J., {Barnes}, J., {et~al.} 2016, \nat, 536,
  437

\bibitem[{{Barnes} {et~al.}(2016){Barnes}, {Deitrick}, {Luger}, {Driscoll},
  {Quinn}, {Fleming}, {Guyer}, {McDonald}, {Meadows}, {Arney}, {Crisp},
  {Domagal-Goldman}, {Lincowski}, {Lustig-Yaeger}, \&
  {Schwieterman}}]{Barnes2016}
{Barnes}, R., {Deitrick}, R., {Luger}, R., {et~al.} 2016, ArXiv e-prints
  [\eprint[arXiv]{1608.06919}]

\bibitem[{{Beuzit} {et~al.}(2008){Beuzit}, {Feldt}, {Dohlen}, {Mouillet},
  {Puget}, {Wildi}, {Abe}, {Antichi}, {Baruffolo}, {Baudoz}, {Boccaletti},
  {Carbillet}, {Charton}, {Claudi}, {Downing}, {Fabron}, {Feautrier},
  {Fedrigo}, {Fusco}, {Gach}, {Gratton}, {Henning}, {Hubin}, {Joos}, {Kasper},
  {Langlois}, {Lenzen}, {Moutou}, {Pavlov}, {Petit}, {Pragt}, {Rabou}, {Rigal},
  {Roelfsema}, {Rousset}, {Saisse}, {Schmid}, {Stadler}, {Thalmann}, {Turatto},
  {Udry}, {Vakili}, \& {Waters}}]{Beuzit2008}
{Beuzit}, J.-L., {Feldt}, M., {Dohlen}, K., {et~al.} 2008, in \procspie, Vol.
  7014, Ground-based and Airborne Instrumentation for Astronomy II, 701418

\bibitem[{{Bottom} {et~al.}(2016){Bottom}, {Shelton}, {Wallace}, {Bartos},
  {Kuhn}, {Mawet}, {Mennesson}, {Burruss}, \& {Serabyn}}]{Bottom2016}
{Bottom}, M., {Shelton}, J.~C., {Wallace}, J.~K., {et~al.} 2016, \pasp, 128,
  075003

\bibitem[{{Boyajian} {et~al.}(2012){Boyajian}, {von Braun}, {van Belle},
  {McAlister}, {ten Brummelaar}, {Kane}, {Muirhead}, {Jones}, {White},
  {Schaefer}, {Ciardi}, {Henry}, {L{\'o}pez-Morales}, {Ridgway}, {Gies}, {Jao},
  {Rojas-Ayala}, {Parks}, {Sturmann}, {Sturmann}, {Turner}, {Farrington},
  {Goldfinger}, \& {Berger}}]{Boyajian2012}
{Boyajian}, T.~S., {von Braun}, K., {van Belle}, G., {et~al.} 2012, \apj, 757,
  112

\bibitem[{{Cahoy} {et~al.}(2010){Cahoy}, {Marley}, \& {Fortney}}]{Cahoy2010}
{Cahoy}, K.~L., {Marley}, M.~S., \& {Fortney}, J.~J. 2010, \apj, 724, 189

\bibitem[{{Chen} \& {Kipping}(2016)}]{Chen2016}
{Chen}, J. \& {Kipping}, D.~M. 2016, ArXiv e-prints
  [\eprint[arXiv]{1603.08614}]

\bibitem[{{Dekker} {et~al.}(2000){Dekker}, {D'Odorico}, {Kaufer}, {Delabre}, \&
  {Kotzlowski}}]{Dekker2000}
{Dekker}, H., {D'Odorico}, S., {Kaufer}, A., {Delabre}, B., \& {Kotzlowski}, H.
  2000, in \procspie, Vol. 4008, Optical and IR Telescope Instrumentation and
  Detectors, ed. M.~{Iye} \& A.~F. {Moorwood}, 534--545

\bibitem[{{Delorme} {et~al.}(2016){Delorme}, {N'Diaye}, {Galicher}, {Dohlen},
  {Baudoz}, {Caillat}, {Rousset}, {Soummer}, \& {Dupuis}}]{Delorme2016}
{Delorme}, J.~R., {N'Diaye}, M., {Galicher}, R., {et~al.} 2016, \aap, 592, A119

\bibitem[{{Demory} {et~al.}(2009){Demory}, {S{\'e}gransan}, {Forveille},
  {Queloz}, {Beuzit}, {Delfosse}, {di Folco}, {Kervella}, {Le Bouquin},
  {Perrier}, {Benisty}, {Duvert}, {Hofmann}, {Lopez}, \& {Petrov}}]{Demory2009}
{Demory}, B.-O., {S{\'e}gransan}, D., {Forveille}, T., {et~al.} 2009, \aap,
  505, 205

\bibitem[{{Dohlen} {et~al.}(2016){Dohlen}, {Vigan}, {Mouillet}, {Wildi},
  {Sauvage}, {Fusco}, {Beuzit}, {Puget}, {Le Mignant}, {Roelfsema}, {Pragt},
  {Schmid}, {Gratton}, {Mesa}, {Claudi}, {Langlois}, {Costille}, {Hugot},
  {O'Neil}, {Guerra}, {N'Diaye}, {Girard}, {Mawet}, \& {Zins}}]{Dohlen2016}
{Dohlen}, K., {Vigan}, A., {Mouillet}, D., {et~al.} 2016, in \procspie, Vol.
  9908, Society of Photo-Optical Instrumentation Engineers (SPIE) Conference
  Series, 99083D

\bibitem[{{Dressing} {et~al.}(2015){Dressing}, {Charbonneau}, {Dumusque},
  {Gettel}, {Pepe}, {Collier Cameron}, {Latham}, {Molinari}, {Udry}, {Affer},
  {Bonomo}, {Buchhave}, {Cosentino}, {Figueira}, {Fiorenzano}, {Harutyunyan},
  {Haywood}, {Johnson}, {Lopez-Morales}, {Lovis}, {Malavolta}, {Mayor},
  {Micela}, {Motalebi}, {Nascimbeni}, {Phillips}, {Piotto}, {Pollacco},
  {Queloz}, {Rice}, {Sasselov}, {S{\'e}gransan}, {Sozzetti}, {Szentgyorgyi}, \&
  {Watson}}]{Dressing2015}
{Dressing}, C.~D., {Charbonneau}, D., {Dumusque}, X., {et~al.} 2015, \apj, 800,
  135

\bibitem[{{Esposito} {et~al.}(2011){Esposito}, {Riccardi}, {Pinna}, {Puglisi},
  {Quir{\'o}s-Pacheco}, {Arcidiacono}, {Xompero}, {Briguglio}, {Agapito},
  {Busoni}, {Fini}, {Argomedo}, {Gherardi}, {Brusa}, {Miller}, {Guerra},
  {Stefanini}, \& {Salinari}}]{Esposito2011}
{Esposito}, S., {Riccardi}, A., {Pinna}, E., {et~al.} 2011, in \procspie, Vol.
  8149, Astronomical Adaptive Optics Systems and Applications IV, 814902

\bibitem[{{Fusco} {et~al.}(2016){Fusco}, {Sauvage}, {Mouillet}, {Costille},
  {Petit}, {Beuzit}, {Dohlen}, {Milli}, {Girard}, {Kasper}, {Vigan}, {Suarez},
  {Soenke}, {Downing}, {N'Diaye}, {Baudoz}, {Sevin}, {Baruffolo}, {Schmid},
  {Salasnich}, {Hugot}, \& {Hubin}}]{Fusco2016}
{Fusco}, T., {Sauvage}, J.-F., {Mouillet}, D., {et~al.} 2016, in \procspie,
  Vol. 9909, Society of Photo-Optical Instrumentation Engineers (SPIE)
  Conference Series, 99090U

\bibitem[{{Gillon} {et~al.}(2016){Gillon}, {Jehin}, {Lederer}, {Delrez}, {de
  Wit}, {Burdanov}, {Van Grootel}, {Burgasser}, {Triaud}, {Opitom}, {Demory},
  {Sahu}, {Bardalez Gagliuffi}, {Magain}, \& {Queloz}}]{Gillon2016}
{Gillon}, M., {Jehin}, E., {Lederer}, S.~M., {et~al.} 2016, \nat, 533, 221

\bibitem[{{Giver}(1978)}]{Giver1978}
{Giver}, L.~P. 1978, \jqsrt, 19, 311

\bibitem[{{Hinkle} {et~al.}(2003){Hinkle}, {Wallace}, \&
  {Livingston}}]{Hinkle2003}
{Hinkle}, K.~H., {Wallace}, L., \& {Livingston}, W. 2003, in Bulletin of the
  American Astronomical Society, Vol.~35, American Astronomical Society Meeting
  Abstracts, 1260

\bibitem[{{Husser} {et~al.}(2013){Husser}, {Wende-von Berg}, {Dreizler},
  {Homeier}, {Reiners}, {Barman}, \& {Hauschildt}}]{Husser2013}
{Husser}, T.-O., {Wende-von Berg}, S., {Dreizler}, S., {et~al.} 2013, \aap,
  553, A6

\bibitem[{{Jones} {et~al.}(2013){Jones}, {Noll}, {Kausch}, {Szyszka}, \&
  {Kimeswenger}}]{Jones2013}
{Jones}, A., {Noll}, S., {Kausch}, W., {Szyszka}, C., \& {Kimeswenger}, S.
  2013, \aap, 560, A91

\bibitem[{{Karkoschka}(1998)}]{Karkoschka1998}
{Karkoschka}, E. 1998, \icarus, 133, 134

\bibitem[{{Kervella} {et~al.}(2016){Kervella}, {Lagadec}, {Montarg{\`e}s},
  {Ridgway}, {Chiavassa}, {Haubois}, {Schmid}, {Langlois}, {Gallenne}, \&
  {Perrin}}]{Kervella2016}
{Kervella}, P., {Lagadec}, E., {Montarg{\`e}s}, M., {et~al.} 2016, \aap, 585,
  A28

\bibitem[{{Kipping} {et~al.}(2016){Kipping}, {Cameron}, {Hartman}, {Davenport},
  {Matthews}, {Sasselov}, {Rowe}, {Siverd}, {Chen}, {Sandford}, {Bakos},
  {Jordan}, {Bayliss}, {Henning}, {Mancini}, {Penev}, {Csubry}, {Bhatti}, {Da
  Silva Bento}, {Guenther}, {Kuschnig}, {Moffat}, {Rucinski}, \&
  {Weiss}}]{Kipping2016}
{Kipping}, D.~M., {Cameron}, C., {Hartman}, J.~D., {et~al.} 2016, ArXiv
  e-prints [\eprint[arXiv]{1609.08718}]

\bibitem[{{Kreidberg} \& {Loeb}(2016)}]{Kreidberg2016}
{Kreidberg}, L. \& {Loeb}, A. 2016, \apjl, 832, L12

\bibitem[{{Leconte} {et~al.}(2015){Leconte}, {Wu}, {Menou}, \&
  {Murray}}]{Leconte2015}
{Leconte}, J., {Wu}, H., {Menou}, K., \& {Murray}, N. 2015, Science, 347, 632

\bibitem[{{Mawet} {et~al.}(2013){Mawet}, {Absil}, {Delacroix}, {Girard},
  {Milli}, {O'Neal}, {Baudoz}, {Boccaletti}, {Bourget}, {Christiaens},
  {Forsberg}, {Gonte}, {Habraken}, {Hanot}, {Karlsson}, {Kasper}, {Lizon},
  {Muzic}, {Olivier}, {Pe{\~n}a}, {Slusarenko}, {Tacconi-Garman}, \&
  {Surdej}}]{Mawet2013}
{Mawet}, D., {Absil}, O., {Delacroix}, C., {et~al.} 2013, \aap, 552, L13

\bibitem[{{Mayor} {et~al.}(2003){Mayor}, {Pepe}, {Queloz}, {Bouchy},
  {Rupprecht}, {Lo Curto}, {Avila}, {Benz}, {Bertaux}, {Bonfils}, {Dall},
  {Dekker}, {Delabre}, {Eckert}, {Fleury}, {Gilliotte}, {Gojak}, {Guzman},
  {Kohler}, {Lizon}, {Longinotti}, {Lovis}, {Megevand}, {Pasquini}, {Reyes},
  {Sivan}, {Sosnowska}, {Soto}, {Udry}, {van Kesteren}, {Weber}, \&
  {Weilenmann}}]{Mayor2003}
{Mayor}, M., {Pepe}, F., {Queloz}, D., {et~al.} 2003, The Messenger, 114, 20

\bibitem[{{Meadows} {et~al.}(2016){Meadows}, {Arney}, {Schwieterman},
  {Lustig-Yaeger}, {Lincowski}, {Robinson}, {Domagal-Goldman}, {Barnes},
  {Fleming}, {Deitrick}, {Luger}, {Driscoll}, {Quinn}, \&
  {Crisp}}]{Meadows2016}
{Meadows}, V.~S., {Arney}, G.~N., {Schwieterman}, E.~W., {et~al.} 2016, ArXiv
  e-prints [\eprint[arXiv]{1608.08620}]

\bibitem[{{Milli} {et~al.}(2013){Milli}, {Mouillet}, {Mawet}, {Schmid},
  {Bazzon}, {Girard}, {Dohlen}, \& {Roelfsema}}]{Milli2013}
{Milli}, J., {Mouillet}, D., {Mawet}, D., {et~al.} 2013, \aap, 556, A64

\bibitem[{{Noll} {et~al.}(2012){Noll}, {Kausch}, {Barden}, {Jones}, {Szyszka},
  {Kimeswenger}, \& {Vinther}}]{Noll2012}
{Noll}, S., {Kausch}, W., {Barden}, M., {et~al.} 2012, \aap, 543, A92

\bibitem[{{Otten} {et~al.}(2014){Otten}, {Snik}, {Kenworthy}, {Miskiewicz}, \&
  {Escuti}}]{Otten2014}
{Otten}, G.~P.~P.~L., {Snik}, F., {Kenworthy}, M.~A., {Miskiewicz}, M.~N., \&
  {Escuti}, M.~J. 2014, Optics Express, 22, 30287

\bibitem[{{Pepe} {et~al.}(2014){Pepe}, {Molaro}, {Cristiani}, {Rebolo},
  {Santos}, {Dekker}, {M{\'e}gevand}, {Zerbi}, {Cabral}, {Di Marcantonio},
  {Abreu}, {Affolter}, {Aliverti}, {Allende Prieto}, {Amate}, {Avila},
  {Baldini}, {Bristow}, {Broeg}, {Cirami}, {Coelho}, {Conconi}, {Coretti},
  {Cupani}, {D'Odorico}, {De Caprio}, {Delabre}, {Dorn}, {Figueira}, {Fragoso},
  {Galeotta}, {Genolet}, {Gomes}, {Gonz{\'a}lez Hern{\'a}ndez}, {Hughes},
  {Iwert}, {Kerber}, {Landoni}, {Lizon}, {Lovis}, {Maire}, {Mannetta},
  {Martins}, {Monteiro}, {Oliveira}, {Poretti}, {Rasilla}, {Riva}, {Santana
  Tschudi}, {Santos}, {Sosnowska}, {Sousa}, {Span{\'o}}, {Tenegi}, {Toso},
  {Vanzella}, {Viel}, \& {Zapatero Osorio}}]{Pepe2014}
{Pepe}, F., {Molaro}, P., {Cristiani}, S., {et~al.} 2014, Astronomische
  Nachrichten, 335, 8

\bibitem[{{Riaud} \& {Schneider}(2007)}]{Riaud2007}
{Riaud}, P. \& {Schneider}, J. 2007, \aap, 469, 355

\bibitem[{{Ribas} {et~al.}(2016){Ribas}, {Bolmont}, {Selsis}, {Reiners},
  {Leconte}, {Raymond}, {Engle}, {Guinan}, {Morin}, {Turbet}, {Forget}, \&
  {Anglada-Escude}}]{Ribas2016}
{Ribas}, I., {Bolmont}, E., {Selsis}, F., {et~al.} 2016, ArXiv e-prints
  [\eprint[arXiv]{1608.06813}]

\bibitem[{{Robinson} {et~al.}(2011){Robinson}, {Meadows}, {Crisp}, {Deming},
  {A'Hearn}, {Charbonneau}, {Livengood}, {Seager}, {Barry}, {Hearty},
  {Hewagama}, {Lisse}, {McFadden}, \& {Wellnitz}}]{Robinson2011}
{Robinson}, T.~D., {Meadows}, V.~S., {Crisp}, D., {et~al.} 2011, Astrobiology,
  11, 393

\bibitem[{{Rodler} \& {L{\'o}pez-Morales}(2014)}]{Rodler2014}
{Rodler}, F. \& {L{\'o}pez-Morales}, M. 2014, \apj, 781, 54

\bibitem[{{Rogers}(2015)}]{Rogers2015}
{Rogers}, L.~A. 2015, \apj, 801, 41

\bibitem[{{Schmid} {et~al.}(2006){Schmid}, {Beuzit}, {Feldt}, {Gisler},
  {Gratton}, {Henning}, {Joos}, {Kasper}, {Lenzen}, {Mouillet}, {Moutou},
  {Quirrenbach}, {Stam}, {Thalmann}, {Tinbergen}, {Verinaud}, {Waters}, \&
  {Wolstencroft}}]{Schmid2006}
{Schmid}, H.~M., {Beuzit}, J.-L., {Feldt}, M., {et~al.} 2006, in IAU Colloq.
  200: Direct Imaging of Exoplanets: Science and Techniques, ed. C.~{Aime} \&
  F.~{Vakili}, 165--170

\bibitem[{{Snellen} {et~al.}(2015){Snellen}, {de Kok}, {Birkby}, {Brandl},
  {Brogi}, {Keller}, {Kenworthy}, {Schwarz}, \& {Stuik}}]{Snellen2015}
{Snellen}, I., {de Kok}, R., {Birkby}, J.~L., {et~al.} 2015, \aap, 576, A59

\bibitem[{{Snellen} {et~al.}(2014){Snellen}, {Brandl}, {de Kok}, {Brogi},
  {Birkby}, \& {Schwarz}}]{Snellen2014}
{Snellen}, I.~A.~G., {Brandl}, B.~R., {de Kok}, R.~J., {et~al.} 2014, \nat,
  509, 63

\bibitem[{{Sparks} \& {Ford}(2002)}]{Sparks2002}
{Sparks}, W.~B. \& {Ford}, H.~C. 2002, \apj, 578, 543

\bibitem[{{Turbet} {et~al.}(2016){Turbet}, {Leconte}, {Selsis}, {Bolmont},
  {Forget}, {Ribas}, {Raymond}, \& {Anglada-Escud{\'e}}}]{Turbet2016}
{Turbet}, M., {Leconte}, J., {Selsis}, F., {et~al.} 2016, ArXiv e-prints
  [\eprint[arXiv]{1608.06827}]

\bibitem[{{Wolfgang} \& {Lopez}(2015)}]{Wolfgang2015}
{Wolfgang}, A. \& {Lopez}, E. 2015, \apj, 806, 183

\end{thebibliography}

\end{document}